\documentclass[sigconf]{acmart}
\usepackage{hyperref}
\usepackage{multirow}
\usepackage{dblfloatfix}
\usepackage{booktabs}
\usepackage{array}
\usepackage{subfigure}
\usepackage[utf8]{inputenc}
\usepackage{algorithm}
\usepackage{algpseudocode}
\usepackage{setspace}
\usepackage{enumitem}
\usepackage{cleveref}
\usepackage{balance} 
\usepackage{mathrsfs}
\usepackage{multirow}
\usepackage{ulem}
\usepackage{makecell} 
\usepackage{graphicx} 
\usepackage{booktabs,multirow,tabularx,threeparttable,xcolor}
\usepackage[table,xcdraw]{xcolor}   

\crefformat{section}{§#2#1#3}
\renewcommand\arraystretch{0.95}
\setlist[itemize]{leftmargin=*, itemsep=0.3em, topsep=0.5em, partopsep=1ex}

\copyrightyear{2026}
\acmYear{2026}
\setcopyright{cc}
\setcctype{by}
\acmConference[KDD '26]{Proceedings of the 32nd ACM SIGKDD Conference on Knowledge Discovery and Data Mining V.1}{August 09--13, 2026}{Jeju Island, Republic of Korea}
\acmBooktitle{Proceedings of the 32nd ACM SIGKDD Conference on Knowledge Discovery and Data Mining V.1 (KDD '26), August 09--13, 2026, Jeju Island, Republic of Korea}
\acmPrice{}
\acmDOI{10.1145/3770854.3783938}
\acmISBN{979-8-4007-2258-5/2026/08}

\AtBeginDocument{%
  }

\begin{document}

\title{\textsc{KLAN}: Kuaishou Landing-page Adaptive Navigator
}



\author{Fan Li}
\authornote{Both authors contributed equally to this research.}
\authornote{Work done during an internship at Kuaishou Technology.}
\affiliation{%
\institution{Duke University}
\city{Durham}
   \country{USA}
 }
 \email{fan.li@duke.edu}

\author{Chang Meng}
\authornotemark[1]
\affiliation{%
   \institution{Kuaishou Technology}
   \city{Beijing}
   \country{China}
 }
 \email{mengchang@kuaishou.com}

 \author{Jiaqi Fu}
 \affiliation{%
   \institution{Kuaishou Technology}
   \city{Beijing}
   \country{China}
 }
 \email{fujiaqi05@kuaishou.com}

 \author{Shuchang Liu}
 \affiliation{%
   \institution{Kuaishou Technology}
   \city{Beijing}
   \country{China}
 }
 \email{liushuchang@kuaishou.com}

 \author{Jiashuo Zhang}
 \authornotemark[2]
 \affiliation{%
   \institution{Kuaishou Technology}
   \city{Beijing}
   \country{China}
 }
 \email{zhangjiashuo03@kuaishou.com}

 \author{Tianke Zhang}
 \affiliation{%
   \institution{Kuaishou Technology}
   \city{Beijing}
   \country{China}
 }
 \email{zhangtianke@kuaishou.com}

 \author{Xueliang Wang}
 \authornote{The corresponding author.}
 \affiliation{%
   \institution{Kuaishou Technology}
   \city{Beijing}
   \country{China}
 }
 \email{wangxueliang03@kuaishou.com}

 \author{Xiaoqiang Feng}
 \authornotemark[3]
 \affiliation{%
   \institution{Kuaishou Technology}
   \city{Beijing}
   \country{China}
 }
 \email{fengxiaoqiang@kuaishou.com}

\renewcommand{\shortauthors}{Fan Li et al.}

\begin{abstract}
Modern online platforms configure multiple pages to accommodate diverse user needs.
This multi-page architecture inherently establishes a two-stage interaction paradigm between the user and the platform: 
\textbf{(1) Stage I: page navigation}, navigating users to a specific page 
and \textbf{(2) Stage II: in-page interaction}, where users engage with customized content within the specific page. 
While the majority of research has been focusing on the sequential recommendation task that improves users' feedback in Stage II, 
there has been little investigation on how to achieve better page navigation in Stage I.
To fill this gap, we formally define the task of Personalized Landing Page Modeling (PLPM) into the field of recommender systems: Given a user upon app entry, the goal of PLPM is to proactively select the most suitable landing page from a set of candidates (e.g., functional tabs, content channels, or aggregation pages) to optimize the short-term metric Page Drop-off Ratio (PDR) and the long-term user engagement and satisfaction metrics, while adhering to industrial constraints.
Additionally, we propose KLAN (\underline{\textbf{K}}uaishou \underline{\textbf{L}}anding-page \underline{\textbf{A}}daptive \underline{\textbf{N}}avigator), a hierarchical solution framework designed to provide personalized landing pages under the formulation of PLPM.
KLAN comprises three key components: \textbf{(1) KLAN-ISP} captures inter-day static page preference; \textbf{(2) KLAN-IIT} captures intra-day dynamic interest transitions and \textbf{(3) KLAN-AM} adaptively integrates both components for optimal navigation decisions.
Extensive online experiments conducted on the Kuaishou platform demonstrate the effectiveness of KLAN, obtaining \textbf{+0.205\%} and \textbf{+0.192\%} improvements on in Daily Active Users (DAU) and user Lifetime (LT).
Our KLAN is ultimately deployed on the
online platform at full traffic, serving hundreds of millions of users.
To promote further research in this important area, we will release our dataset and code upon paper acceptance.
\end{abstract}




\maketitle

\begin{figure}[t]
\includegraphics[width=0.93\linewidth]{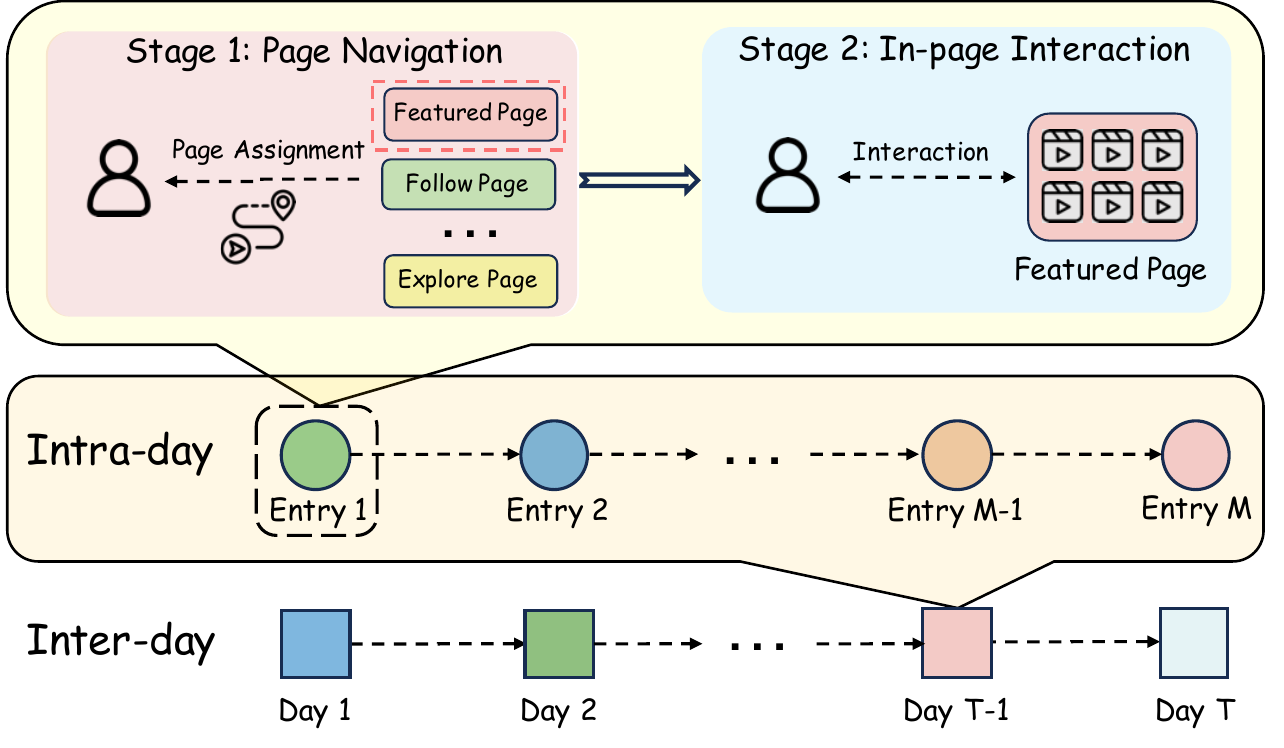}
\caption{An example of the two-stage interaction paradigm on multi-page platforms: the user is first navigated to a functional or channel page, and then interacts with content within that page.}
\label{fig:twostage}
\vspace{-0.4cm}
\end{figure}
\begin{figure*}[!t]
    \centering  \includegraphics[width=1.00\linewidth]{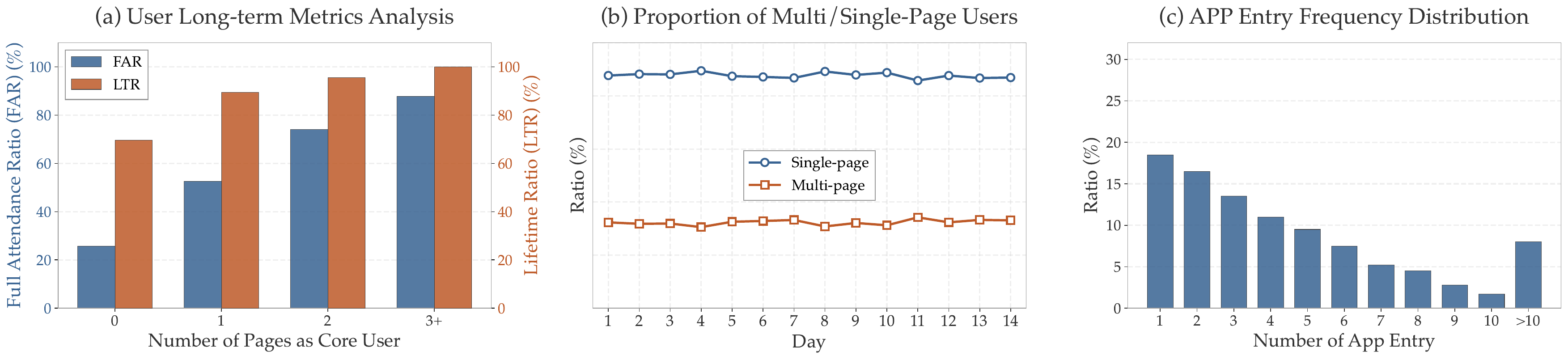}
    \caption{(a) Users' long-term metric performance across different active page counts, where core users are those whose time spent on a page exceeds a certain percentile. Full attendance ratio denotes the proportion of the users who are active each day within the statistical period and LT is an unbiased estimate of medium and long-term DAU for experimental evaluation (as detailed in Equation \ref{equ:LT7}). Here we use 7-day Lifetime and LTR is calculated with a normalization operation ($\frac{\textbf{LT}}{\textbf{7}}$); (b) The proportion of multi/single-page users over 14 consecutive days; (c) The distribution of users based on their number of app entry within a single day. } 
    \label{fig:DA}
    \vspace{-0.25cm}
\end{figure*}   

\section{Introduction}

With the prevalence of Web 2.0 and mobile devices, 
a massive number of users have joined online platforms such as Amazon, TikTok, and Kuaishou for shopping, content creation, and social networking.
On these platforms, multiple functional and channel pages are configured to cater to users' diverse needs.
For example, Kuaishou offers various types of pages including the "Featured" page that presents recent and personalized recommendations to users and the "Explore" page that provides less personalized but more diverse video recommendations.
Inside each of these pages, users interact with the recommender system with consecutive requests, receiving sequential recommendations and providing feedback~\cite{PROWTP,CVRDD,ABREC,zhai2025combinatorialoptimizationperspectivebased,zhai2025heterogeneousmultitreatmentupliftmodeling}.
In this sense, we consider the multi-page online recommender system as a two-stage interaction paradigm between the user and the platform (shown in Figure \ref{fig:twostage}): an initial \textbf{page navigation} stage that navigates users to a specific functional or channel page, and a subsequent \textbf{in-page interaction} stage where users sequentially engage with customized content on the specific page.


While most existing approaches for video recommendation platforms focus on the sequential recommendation task in the second in-page interaction stage~\cite{kang2018self,sun2019bert4rec,mao2025distinguished,mao2024invariant,       guo2023compressed,meng2023parallel,meng2023hierarchical,meng2024coarsetofinedynamicupliftmodeling,chang2023twin,pi2020search,CDM_Fan,CPF_Fan,yu2025mattersbridgingtopicssocial}, there has been little investigation on how to achieve better page navigation in the first stage. 
However, the initial page navigation stage significantly impacts the user experience in the system.
Firstly, it determines whether the following interactions can occur and directly influences the user's first impression of the online service, which is reflected by short-term metrics like the Page Drop-off Ratio (PDR), as well as the long-term metrics such as Daily Active Users (DAU) and LifeTime (LT).
Here, we define the \textbf{Page Drop-off Ratio (PDR)} as the probability that a user immediately leaves the assigned initial page upon app entry.
Furthermore, users may typically enter the app multiple times in one day (Figure \ref{fig:DA}-c), so effective page navigation may choose different initial pages that accurately address the users' dynamic demands upon each entry.
Yet, inaccurate navigation could cause the users to frequently switch pages (i.e. the page drop-off), so it is also critical to avoid this negative impact on the user experience.
Additionally, empirical evidence from Kuaishou, as shown in Figure \ref{fig:DA}-a and \ref{fig:DA}-b, demonstrates that approximately 42\% of users engage across multiple pages and consistently outperform single-page users in terms of DAU and LT.
While the remaining 58\% of users exhibit single-page activities within a single day (shown in Figure \ref{fig:DA}-b),
we believe that when the system properly introduces new pages to these users, it is possible to stimulate their potential in multi-page consumption, which improves the exposure of diverse video contents on different pages and eventually motivates the sustainable advancement of the platform's ecosystem.

To this end, we formally define the task of \textbf{Personalized Landing Page Modeling (PLPM)} as the following decision-making problem: Given a user upon app entry, the goal of PLPM is to proactively select the most suitable landing page from a set of candidates (e.g., functional tabs, content channels, or aggregation pages) to optimize the short-term PDR metric and the long-term user engagement and satisfaction metrics, while adhering to industrial constraints.
Despite the demonstrated importance of this task, current industrial practices remain surprisingly rudimentary. 
For example, page navigation decisions are often made using simple heuristics, such as resuming the page from which the user last exited, or selecting the most frequently visited page based on offline historical statistics, etc.
These approaches lack real-time adaptability and fine-grained personalization, and are thus prone to misjudging users’ current intents.
Consequently, there is an urgent need to develop a new modeling paradigm to make more accurate page navigation decisions, improving user experience and retention, while adhering to industrial constraints such as effective page entry frequency and latency requirements.



To effectively address the PLPM task, a deep understanding of user behavior upon app entry is essential. 
On short-video platforms, user behavior exhibits a dual nature~\cite{Zhu_2024,zhu2025longtermclockfinegrainedtime,meng2024coarsetofinedynamicupliftmodeling}, shaped by both static preferences and dynamic interest transitions.
As shown in Figure \ref{fig:DA}-b, we also observe this dual pattern in Kuaishou's short-video platform as users may display intra-day page shifts or keep single-page interactions in one day.
Nevertheless, users with daily single-page interactions may also actively switch pages across days.
Therefore, an effective solution for PLPM necessitates a hierarchical model each poses different challenges:
\begin{itemize}
    \item \textbf{Inter-day static page preferences modeling}: 
    User's static page preference expresses the long-term stable preferences towards each pages, which is a critical feature to model the landing page of a user's first entry on each day.
    Empirically, users have imbalanced interactions across pages, which introduces considerable exposure bias.
    For example, we observe that the vast majority of samples are available on the "Featured" page.
    On the other hand, different pages are heterogeneous in presentation format, including content and interaction patterns, which results in vastly different user response distributions across pages (Section \ref{sec:page_heterogeneity}).
    Given such heterogeneity, modeling user engagement for each page in isolation fails to capture the underlying causal relationships. Instead, a holistic perspective is needed—one that explicitly accounts for how assigning a user to one page over another influences their subsequent behavior. While uplift modeling provides a principled framework for estimating such treatment effects, existing approaches often rely on joint optimization over multiple objectives, which introduces gradient conflict problem, ultimately leading to suboptimal performance in practice.
    
    %

    \item \textbf{Intra-day dynamic interest transition modeling}:
    As shown in Figure \ref{fig:DA}-c, 70\%+ of users open the app more than once within a single day. 
    This necessitates delivering appropriate landing pages multiple times throughout the day, which requires a personalized model tailored to each user's evolving preferences.
    Yet, this is inherently a challenging task, as users' dynamic interests are highly sensitive to immediate contextual triggers (e.g., live-stream notifications), resulting in highly volatile behaviors.
    Moreover, the exposure bias of landing pages also exists in intra-day data, which further amplifies the challenge.
    Furthermore, intra-day dynamic interests and inter-day static page preferences reflect fundamentally different behavioral patterns, which may require fine-grained balancing strategy in real-time.
    

    

    
\end{itemize}






To address the aforementioned challenges, we propose KLAN (\underline{\textbf{K}}uaishou \underline{\textbf{L}}anding-page \underline{\textbf{A}}daptive \underline{\textbf{N}}avigator), a hierarchical solution framework designed to provide personalized landing pages under the formulation of PLPM.
Specifically, we first introduce the Inter-day Static Page Preferences module (KLAN-ISP), which models users' page uplift across different pages to capture their static preferences for each corresponding page. 
It adopts causal modeling within each page respectively based on our carefully constructed datasets, effectively addressing the challenge of page heterogeneity and exposure bias.
Then, we develop the Intra-day Interest Transition module (KLAN-IIT), which leverages reinforcement learning algorithms to capture users' dynamic interest transitions in real-time. 
To mitigate exposure bias within Intra-day datasets, we adopt Conservative Q-Learning to enable stable policy learning. 
Furthermore, we enhance CQL with a Dynamic Conservative Coefficient that automatically optimizes the exploration-exploitation balance, ensuring both stability and adaptability in policy updates.
Finally, to integrate inter-day and intra-day modeling strategies, we introduce an Adaptive Modulation module (KLAN-AM), which monitors the relative contributions of both components, enabling adaptive decision-making for personalized landing page navigation.

Our contributions can be summarized as follows:
\begin{itemize}
    \item To the best of our knowledge, we are the first to formally introduce the PLPM problem in recommender system, providing systematic analysis of its importance and associated technical challenges in the hierarchical model.
    \item We propose KLAN, a pioneering comprehensive framework specifically designed for PLPM.
    \item We conducted extensive online experiments on the Kuaishou platform, obtaining \textbf{+0.205\%} and \textbf{+0.192\%} improvements on DAU and LT respectively. Our KLAN is ultimately deployed on the online platform at full traffic, serving hundreds of millions of users.
    \item Our work fills the research gap in personalized landing page modeling. To facilitate further development and advancement in this field, we will release both the datasets and the code after the acceptance of this paper.
\end{itemize}

\section{Related Work}
\subsection{Uplift Model for Video Recommendation Platform}
Uplift models aim to identify the target user groups for each specific treatment by estimating individual treatment effects (ITE). Existing approaches can be categorized into three main methodologies: Meta-learner based methods, Tree-based methods and Neural Networks based methods.
Meta-learner based methods~\cite{kunzel2019metalearners, yao2021survey} constitute a classical framework for uplift modeling.
The S-learner~\cite{kunzel2019metalearners} treats the treatment variable as a feature within a single model, while the T-learner~\cite{kunzel2019metalearners} builds separate models for treatment and control groups.
Tree-based methods~\cite{athey2016recursive, nandy2023generalized, radcliffe2011real, wang2015robust} utilize specific tree or forest structures, setting direct computational methods for incremental uplift modeling.
Due to the ability to capture complex treatment-response relationships and estimate more accurate ITE, neural networks have emerged as a popular framework for uplift modeling.
Existing Neural Network-based uplift methods~\cite{bica2020estimating, louizos2017causal, shi2019adapting, wei2024multitreatment, zhou2023direct, ke2021addressing,zhong2022descn} can be categorized into two categories: joint modeling of treatment and modeling of specific treatment causal effects.
The former estimates the non-treatment score through features and incorporates the treatment feature into the treatment modeling based on this score.
For instance, 
EFIN~\cite{liu2023efin} captures the feature subset related to the control group through explicit feature selfinteraction, then models the sensitivity of non-treatment features to specific treatments using a treatment-aware interaction module, and enhances the model’s robustness through an intervention constraint module.
The latter focuses on modeling the causal effects of specific treatment.
CDUM~\cite{meng2024coarsetofinedynamicupliftmodeling} expands the treatment representation for guidance and indicator within a multi-task framework and proposes an online real-time feature learning network based on video recommendation scenarios.
However, existing methods tend to either overemphasize the collaborative effects or the causal effects, making it difficult to adapt to short video recommendation scenarios which involves heterogeneous multi-treatment.

\subsection{Reinforcement Learning for Video Recommendation Platform}

Unlike traditional methods optimizing for the one-step recommendation, reinforcement learning~\cite{sutton1998reinforcement} approaches model video recommendation as a Markov Decision Process, optimizing cumulative rewards to maximize long-term user satisfaction. 
Reinforcement learning (RL) approaches can be broadly categorized into on-policy~\cite{rummery1994on,mnih2016asynchronous} and off-policy~\cite{haarnoja2018soft,watkins1989learning,vanhasselt2015deepreinforcementlearningdouble} methods.
Due to the high policy variance and low sample efficiency of on-policy methods, off-policy methods have emerged as the predominant choice in industrial application.
Among off-policy methods, policy gradient approaches~\cite{chen2019large, chen2019topk, ge2021towards, ge2022toward, li2022autolossgen, xian2019reinforcement} suffer from the mismatch between the behavior policy and the target policy, leading to substantial discrepancies between offline evaluation and online performance.
Value gradient approaches~\cite{vanhasselt2015deepreinforcementlearningdouble,pei2019value, taghipour2007usage, zhao2021dear, zhao2018recommendations}, prominently represented by DQN~\cite{vanhasselt2015deepreinforcementlearningdouble}, have demonstrated superior stability and data efficiency by leveraging experience replay.
However, the traditional DQN struggles with the overestimation bias due to the exposure imbalance in our proposed PLPM scenario.
To address it, conservative reinforcement learning algorithms such as CQL~\cite{kumar2020conservativeqlearningofflinereinforcement} and IQL~\cite{kostrikov2021offlinereinforcementlearningimplicit} incorporate pessimistic value estimation to handle distributional shift.
Despite the theoretical advantages of CQL, its practical deployment reveals a critical hyperparameter sensitivity issue. 
Specifically, an excessively large conservative coefficient leads to overly pessimistic policies that fail to stimulate multi-page user engagement, while an insufficient coefficient approximates standard DQN behavior, resulting in policy instability that violates online serving constraints and degrades user experience. 
Furthermore, our empirical analysis on Kuaishou reveals that intra-day RL decision exhibit a "tidal phenomenon" (detailed in Appendix \ref{sec:TIDAL}).
Motivated by this, we introduce a Dynamic Conservative Coefficient, leveraging intra-day real-time features strongly correlated with the "tidal phenomenon" (such as request hour and daily app entry frequency) to automatically balance the exploration-exploitation trade-off.

\section{The PLPM Problem Formulation}

We define the three sub-tasks of PLPM problem as the inter-day modeling of user's static page preferences, the intra-day modeling of user's dynamic page interest transition and the adaptive modulation of both components.

\subsection{The Inter-day Modeling Task}
Following uplift modeling principles, we first conduct a Randomized Controlled Trial (RCT) in a real-world setting to collect user response data under different treatments.
The collected dataset is denoted as $\mathcal{Z} = \{z_1, z_2, \dots, z_N\}$ and $N$ is the number of all instances. Each instance $z_i = (\mathbf{x}_i, \mathbf{t}_i, y_i) $ comprises the user feature $\mathbf{x}_i$, the specific treatment $\mathbf{t}_i$, and the observed response $y_{i}$ (e.g., users' app usage time).
Following the Neyman-Rubin potential outcome framework~\cite{rubin2005causal}, we denote $y_{i}^{\mathbf{t}_i}$ as the response when the user in the $i$-th instance receives a specific treatment $\mathbf{t}_i \in [1, K]$, and $y^{\mathbf{t}_0}_{i}$ as the response under no treatment.
As we can not observe $y_i^{\mathbf{t}_i}$ and $y_i^{\mathbf{t}_0}$ at the same time, there is no true uplift result for each instance, which is known as a key reason uplift modeling differs from traditional supervised learning. 
Thus, we need uplift modeling to estimate the expected individual treatment effect $\tau^k(\mathbf{x}_i) = \mathbb{E}[y_{i}^k \mid \mathbf{t}_i = k, \mathbf{x}_i] - \mathbb{E}[y_{i}^0 \mid \mathbf{t}_i=0, \mathbf{x}_i], k\in [1,K]$.

To address page heterogeneity and exposure bias, we design different treatment-specific branches to independently estimate $\tau^k(\mathbf{x}_i)$, enabling accurate causal effect modeling for each treatment.
Additionally, we adopt a multi-task learning framework to capture cross-treatment synergies, employing KL divergence to constrain the representations of responses induced by non-treatment across branches. This architecture yields treatment-specific outcome pairs from each branch, i.e., $\left\{ \left( \mathbb{E}(y_{i}^{k} \mid \mathbf{t}_i = k, \mathbf{x}_i), \mathbb{E}(y_{i}^{0,k} \mid \mathbf{t}_i = 0, \mathbf{x}_i) \right) \right\}_{k \in [1,K]}
$, for precise uplift estimation.


\subsection{The Intra-day Modeling Task}
\label{sec:klan-iit-def}
We formulate the intra-day interest transition modeling as a Markov Decision Process (MDP) $(\mathcal{S},\mathcal{A},r,\mathcal{P},\gamma)$, defined as followed:
\begin{itemize}
    \item $\mathcal{S}$: The continuous representation space of the user state. Each state $s_t$ is composed of the real-time contextual features and the historical statistical features at step $t$.
    \item $\mathcal{A}$: The action space comprising all available landing page assignments. Each action $a_t$ represents the assignment of a specific landing page upon app entry at step $t$.
    \item $\mathcal{P}$: $\mathcal{S} \times \mathcal{A} \rightarrow \mathcal{S}$, the  state transition function. 
    \item $r(s_t,a_t)$: The immediate reward that captures the user feedback for the landing-page assignment $a_t \in \mathcal{A}$ on user state $s_t \in \mathcal{S}$. Here, we define the app usage duration of the session (i.e., the sequence of interactions that starts with the display of a landing page when the user enters the app and ends when the user exits) as the immediate reward.
    \item $\gamma \in [0,1]$: The discount factor balancing immediate and future rewards.
\end{itemize}
The goal of this MDP is to learn an optimal policy $\pi^{*}$:$\mathcal{S} \rightarrow \mathcal{A}$ that maximizes the expected discounted return:
\begin{equation}
\setlength{\abovedisplayskip}{0.5pt}
\setlength{\belowdisplayskip}{0.5pt}
    \pi^* = \arg\max_{\pi} \mathbb{E}_{\pi} \left[ \sum_{t=0}^{\infty} \gamma^t r(s_t, a_t) \right].
\end{equation}

\subsection{The Adaptive Modulation Task}
\label{sec:def_AM}
In the online video recommendation pipeline, we design a real-time data streaming to model dynamic user behavior tendencies. Each instance is represented as a tuple $ (\mathbf{c}_t,\mathbf{v}_t, \mathbf{k}_t,\boldsymbol{\gamma}_t^{k_t})$, where $\mathbf{c}_t$ denotes the contextual features, $\mathbf{v}_t$ represents the users' long-term behavioral statistics, $\mathbf{k}_t$ indicates the landing page assigned by the platform and $\boldsymbol{\gamma}_t^{k_t}$ is the observed user response at time t (detailed in Section \ref{sec:real_time_streaming}).
Our objective is to leverage both contextual features $\mathbf{c}_t$ and long-term behavioral statistics $\mathbf{v}_t$ to accurately predict users' immediate behavioral tendencies when presented with landing page $\mathbf{k}_t$.
The predicted immediate behavior tendencies serve as an indicator of the temporal dynamics in user behavior, which subsequently informs the adaptive weighting mechanism between static preferences and dynamic interests in our adaptive fusion.



\begin{figure*}
    \centering
    \includegraphics[width=0.88\linewidth]{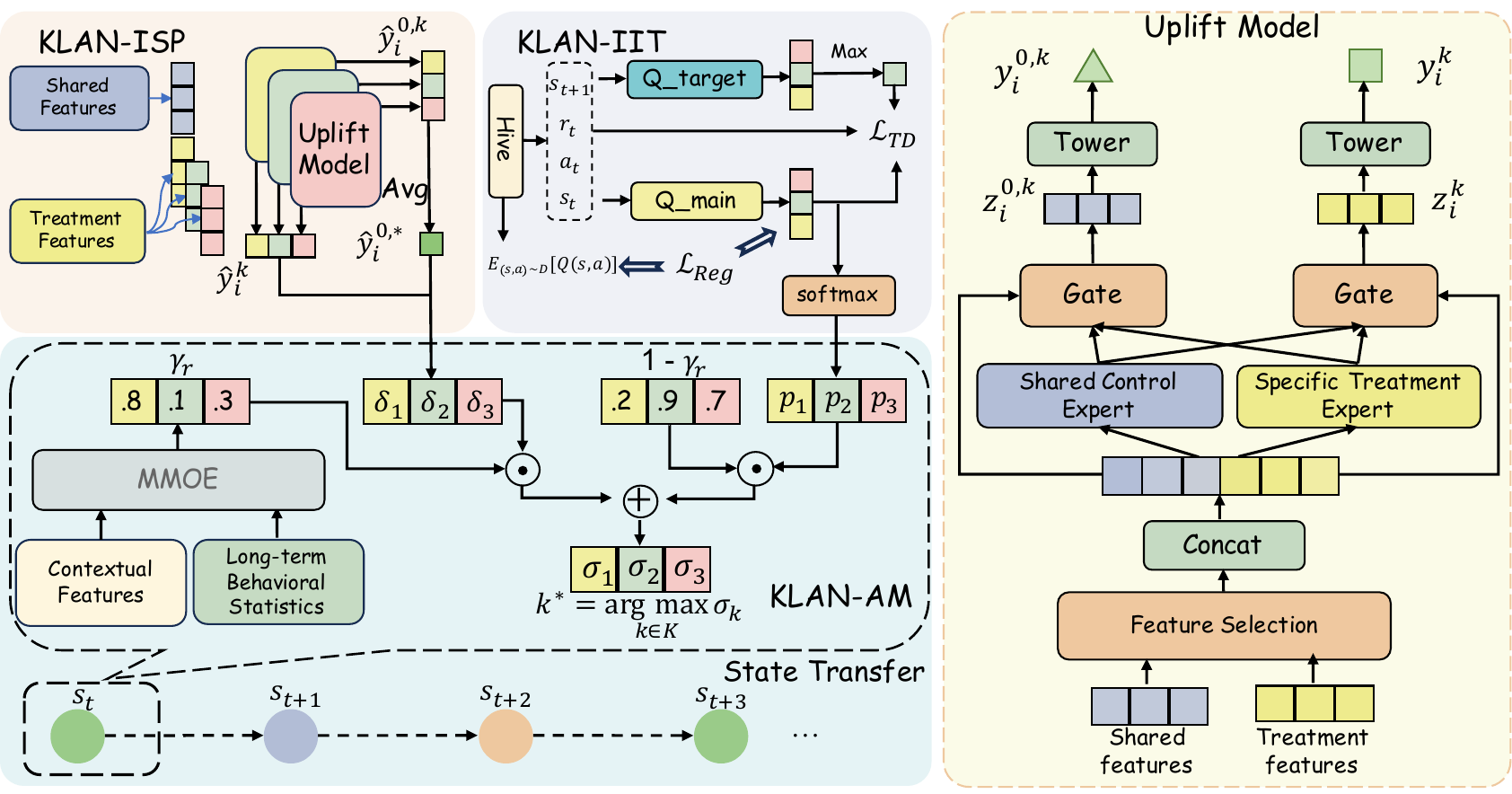}
    \vspace{-0.4cm}
    \caption{The architecture of our proposed method.}
    \label{fig:model}
    \vspace{-0.4cm}
\end{figure*}

\section{Method}
Our KLAN comprises three mudules: \textbf{(1) KLAN-ISP}, for inter-day static page preference modeling; \textbf{(2) KLAN-IIT}, for intra-day dynamic interest transition modeling; \textbf{(3) KLAN-AM}, for adaptive decision-making by integrating both static and dynamic signals.

\subsection{KLAN-ISP: Inter-day Static Preference Modeling}
\label{sec:KLAN-ISP}

\subsubsection{Treatment-Specific Uplift Modeling}

Given an instance $z_i = (\mathbf{x}_i, \mathbf{t}_i, y_i)$ from dataset $\mathcal{Z}$, we construct $K$ treatment-specific branches to independently model the causal effect of each landing page. For the $k$-th branch, the forward computation proceeds as follows:

\textbf{Feature Selection and Fusion:}
First of all, we use trainable embedding layers to obtain the representations of non-treatment features $\mathbf{x}_i$ and treatment features $\mathbf{t}_i$, denoted as $e_{xi}$ and $e_{ti}$, respectively.
We then employ a target attention mechanism for treatment-specific feature selection, extracting relevant features from $\mathbf{e}_{xi}$ guided by $\mathbf{e}_{ti}$
\begin{equation}
\setlength{\abovedisplayskip}{0.5pt}
\setlength{\belowdisplayskip}{0.5pt}
\mathbf{e}_{xi}^k = \text{FS}_k(\mathbf{e}_{xi}, \mathbf{e}_{ti}),
\end{equation}
where $\text{FS}_k(\cdot)$ denotes the $k$-th treatment-specific feature selection module (detailed in \ref{sec:feature_selection}).
These embeddings are then concatenated to form $\mathbf{f}_i^k$, which serves as the input feature vector for the subsequent stage:
\begin{equation}
\setlength{\abovedisplayskip}{0.5pt}
\setlength{\belowdisplayskip}{0.5pt}
    \mathbf{f}_i^k = \text{Concat}(\mathbf{e}_{xi}^k, \mathbf{e}_{ti})
\end{equation}

\textbf{Individual Causal Effect Branches:}
To efficiently model the individual causal effects of each treatment, we design different treatment-specific branches. 
Each treatment-specific branch takes the corresponding feature vector $\mathbf{f}_i^k$ as input and outputs two predictions:$\mathbb{E}(y_{i}^{k} \mid \mathbf{t}_i = k, \mathbf{x}_i) $ and $ \mathbb{E}(y_{i}^{0,k} \mid \mathbf{t}_i = 0, \mathbf{x}_i)$, representing the estimated responses under treatment $k$ and no treatment, respectively.

For the $k$-th branch, we use operations (e.g. MLP) to generate $M$ expert representations $\{\mathbf{d_{im}^k}\}_{m=1}^M$ from $\mathbf{f}_i^k$.
Following the classical operations of multi-task learning, we employ a gating mechanism to aggregate the expert information. 
The final score is then output through a specific tower, using this accumulated information as input, which can be formulated as:
\begin{equation}
\begin{cases}
    \mathbf{g}_i^k &= \text{Softmax}(\mathbf{W}_g^k \cdot \mathbf{f}_i^k + \mathbf{b}_g^k)\\
    \mathbf{z}_{i}^k &=  \sum_{m=1}^{M} \mathbf{g}_i^k(m) \cdot \mathbf{d_{im}^{k}} \\
    \hat{y}_{i}^k &= h^k \left( \mathbf{z}_i^k\right)
\end{cases}
\end{equation}
where $W_g^k \in \mathbb{R}^{M \times d}$ and $b_g^k \in \mathbb{R}^{M \times 1}$ are transformation matrix and bias matrix, and $g_i^k \in \mathbb{R}^{M \times 1}$ represents the gating weights with $\mathbf{g}_i^k(m)$ denoting its $m$-th.element.
$z_i^k$ is the output latent representation of corresponding gate and $h^k(\cdot)$ are the tower functions.

Furthermore, in the case of $k = 0$, the process is similar across branches. However, each branch produces distinct gate outputs and predicted values at this stage. To provide clarity, for a specific branch $k'$, we denote $\mathbf{z}_i^{0,k'}(k' \neq 0)$ and $\hat{y}_i^{0,k'} (k' \neq 0) $ as the individual gate output and predicted value for the $k = 0$ scenario in that branch.

In conclusion, each branch produces two outputs: $\hat{y}_i^k$ (predicted response under treatment $k$) and $\hat{y}_i^{0,k}$ (predicted response under no treatment, estimated by branch $k$).

\subsubsection{Multi-treatment Joint Optimization}
Through Treatment-Specific Uplift Model, we obtain the estimated output pairs $[\hat{y}_i^k,\hat{y}_i^{0,k}]$ and the latent representation pairs $[\mathbf{z}_i^{k},\mathbf{z}_i^{0,k}]$ from each branch.
For training, we adopt the mean square error as the main training loss. 
To improve the robustness and effectiveness of ISP, we apply KL divergence to constrain non-treatment representations across branches.
Therefore, the total training loss for ISP is as follows:
\begin{equation}
\setlength{\abovedisplayskip}{0.5pt}
\setlength{\belowdisplayskip}{0.5pt}
\mathcal{L} = 
\begin{cases}
\frac{1}{N} \sum_{i=1}^{N} \sum_{k'=1}^{K} \left( \left( y_{i}^{k} - \hat{y}_{i}^{0, k'} \right)^2 + D_{\mathrm{KL}}\left( z_{i}^{0, k'} \,\middle\|\, {z}_i^{0,*} \right) \right), & k = 0 \\
\frac{1}{N} \sum_{i=1}^{N} \left( y_{i}^{k} - \hat{y}_{i}^{k} \right)^2, & k \ne 0
\end{cases}
\label{eq:ISP_loss}
\end{equation}
where $y_i^k$ is the response label under $k$-th treatment and ${z}_i^{0,*} = \frac{1}{K}\sum_{k=1}^{K} z_i^{0,k}$.
It is worth noting that in the training stage, since each sample $(\mathbf{x_i},\mathbf{t_i},y_i)$ is obtained by recording response only under the action of a specific treatment (e.g, $k$-th treatment), we only use the output $\hat{y}_{i}^{k}\left( k \neq 0 \right)$ and $\hat{y}_{i}^{0,k'} \ (k' \in [1,K])$
for training, while mask operation is applied to the calculation of loss to other towers.

\subsubsection{Prediction}
During inference, we iteratively replace the treatment feature $\mathbf{t}_i$ to obtain predictions for all treatments: $\{\hat{y}_{i}^{k'},\hat{y}_{i}^{0,k'}\}_{k'=0}^K$.

To improve the accuracy of the user response estimate, we use the average value $\frac{\sum_{k' = 1}^{K} \hat{y}_{i}^{0,k'}}{K}
$ as the adjusted approximate value for the final estimated response without treatment, denoted as $\hat{y}_i^{0,*}$.

The static preference score for treatment $\mathbf{t_k}$ is then calculated as:
\begin{equation}
\setlength{\abovedisplayskip}{0.5pt}
\setlength{\belowdisplayskip}{0.5pt}
    \delta_k(x_i) =
    \frac{\exp(\hat{y}_{i}^k/\hat{y}_{i}^{0,*})}{\sum_{j=1}^{K} \exp (\hat{y}_{i}^j / \hat{y}_{i}^{0,*})}
\end{equation}

By repeating this calculation for each of the $K$ treatments, we obtain the complete set of static preference scores $\{\delta_k(\mathbf{x_i})\}_{k=1}^K$, which are subsequently fed into the Adaptive Module.

\subsection{KLAN-IIT: Intra-day Interest Transition Modeling}
\subsubsection{Conservative Q-Learning for Policy Optimization}

To solve the MDP formulated in Section~\ref{sec:klan-iit-def}, we employ Conservative Q-Learning (CQL), which addresses the challenge of offline reinforcement learning by preventing overestimation of out-of-distribution actions.
We parameterize two Q-networks:(1) Main Q-network $Q^{\pi} : \mathcal{S} \times \mathcal{A} \rightarrow \mathbb{R}^{K\times1}$ with parameters $\theta^{\pi}$; (2) Target Q-network $Q^{\mu}:\mathcal{S} \times \mathcal{A} \rightarrow \mathbb{R}^{K\times1}$ with parameters $\theta^{\mu}$.

The optimal Q-function $Q^*(s,a)$ satisfies the Bellman optimality equation:
\begin{equation}
Q^*(s_t, a_t) = \mathbb{E}_{s_{t+1} \sim \mathcal{P}(\cdot|s_t, a_t)} \left[ r(s_t, a_t) + \gamma \max_{a'} Q^*(s_{t+1}, a') \right]
\end{equation}

\subsubsection{Training Objective}

The CQL algorithm optimizes a regularized objective that combines temporal difference (TD) learning with a conservative penalty:

\textbf{Temporal Difference Loss:}
\begin{equation}
    \mathcal{L}_{TD} = \mathbb{E}_{(s_t, a_t, r_t, s_{t+1}) \sim \mathcal{D}} \left[ \left( Q^{\pi}(s_t, a_t) - \tilde{Q}_t \right)^2 \right]
\end{equation}
where $\tilde{Q}_t = r(s_t, a_t) + \gamma \max_{a'} Q^{\mu}(s_{t+1}, a')$ represents the TD target.

\textbf{Conservative Regularization:}
\begin{equation}
\mathcal{L}_{\textit{Reg}} = \mathbb{E}_{s \sim \mathcal{D}} \left[ \log \sum_{a \in \mathcal{A}} \exp(Q^{\pi}(s, a)) \right] - \mathbb{E}_{(s, a) \sim \mathcal{D}} \left[ Q^{\pi}(s, a) \right]
\end{equation}

This regularization term penalizes Q-values for out-of-distribution actions while maintaining high values for in-distribution state-action pairs.

\textbf{Overall CQL Objective:}
\begin{equation}
\mathcal{L}_{\textit{CQL}} = \mathcal{L}_{\textit{TD}} + \alpha_t \cdot \mathcal{L}_{\textit{Reg}}
\end{equation}
where $\alpha_t > 0$ is the \textbf{Dynamic Conservative Coefficient} (detailed in APPENDIX \ref{sec:TIDAL}), balancing the exploration-exploitation trade-off based on real-time features.

\subsubsection{Prediction}
After training, inspired by softmax, the intra-day interest score for landing page $k$ is computed as:
\begin{equation}
p_k(s_t) = \frac{\exp(Q^{\pi}(s_t, a_k))}{\sum_{j=1}^{K} \exp(Q^{\pi}(s_t, a_j) )}
\end{equation}
Eventually, we can get the intra-day interest scores $\{p_{k}(s_t)\}_{k=1}^K$, serving as input to the Adaptive Module.

\subsection{KLAN-AM: Adaptive Modulation}
As discussed, through KLAN-ISP and KLAN-IIT, we obtain static preference scores $\{\delta_{k}(\mathbf{x_i})\}_{k=1}^K$ and dynamic interest scores $\{p_{k}(s_t)\}_{k=1}^K$, respectively. However, the relative importance of static preferences and dynamic interests varies significantly across different contexts and user states. To address this challenge, we introduce the Adaptive Modulation module, which dynamically balances the contributions of both signals based on real-time data streaming.

\subsubsection{Context-Aware Weight Estimation}
To monitor the relative contributions of inter-day static page preference and intra-day dynamic interest transition, we employ a MMOE~\cite{ma2018modeling} architecture (detailed in \ref{sec:AM_training}) to predict the users’ immediate
behavioral tendencies:
\begin{equation}
\boldsymbol{\gamma}_t = \text{MMOE}(\mathbf{c}_t, \mathbf{v}_t), 
\end{equation}
where $\boldsymbol{\gamma}_t = [\boldsymbol{\gamma}_t^1,\boldsymbol{\gamma}_t^2,\dots,\boldsymbol{\gamma}_t^K]\in [0,1]^K$ represents the context-aware weight vector,  with $\boldsymbol{\gamma}_t^k$ indicating the relative importance of static preferences for landing page $k$ in the current context.

\subsubsection{Adaptive Score Fusion}
The final navigation score for each landing page $k$ is computed by adaptively combining static and dynamic scores:
\begin{equation}
\sigma_k = \gamma_t^k \cdot \delta_k(\mathbf{x}_i) + (1 - \gamma_t^k) \cdot p_k(s_t)
\end{equation}

\subsubsection{Page Assignment}
The optimal landing page is selected based on the highest fused score:
\begin{equation}
k^* = \arg\max_{k \in [1,K]} \sigma_k
\end{equation}
Through this hierarchical architecture, KLAN effectively navigates users to the most appropriate landing page by considering both their inter-day static preferences and intra-day immediate interests, with an adaptive module balancing them in real time.






\section{Experiment}
\label{sec:Exp}

\subsection{Dataset}

To capture both static preferences and dynamic interests, we train KLAN-ISP and KLAN-IIT on daily and hourly datasets, respectively, while KLAN-AM leverages streaming data to learn real-time adaptive fusion scores.

\subsubsection{\textbf{Daily-level Data.}}
\label{sec:RCT}
We construct a large-scale industrial dataset with daily updates, tailored for uplift modeling and causal inference in personalized landing page scenarios. To accurately capture the causal relationship between landing pages and objections, this dataset is built upon unbiased experimental data collected through a carefully designed randomized controlled trial (RCT), in which users were randomly assigned to control and treatment groups. Each instance in the dataset comprises a rich set of features, including: (1) user profile attributes such as age, region, and activity level; (2) general consumption behavior over the past 7 and 30 days, including app usage duration and the number of videos watched; and (3) page-specific engagement features over the past 7 and 14 days, such as page stay duration, page video playtime, and page livestream playtime. In addition, we compute the average app usage time during the experiment period as the response/label. 
The dataset comprises approximately 5.87 million instances, providing sufficient scale for training and evaluating industrial-grade uplift models. We partition the data into training, validation/test sets following an 8:2 ratio, ensuring robust model development and fair performance assessment. This dataset serves as a valuable benchmark for causal modeling in PLPM, featuring high-quality input features and reliable treatment assignments derived from randomized experimentation.

\begin{figure}[t]
	\centering
	\setlength{\belowcaptionskip}{-0.0cm}
	\setlength{\abovecaptionskip}{-0.0cm}
	\includegraphics[width=\linewidth]{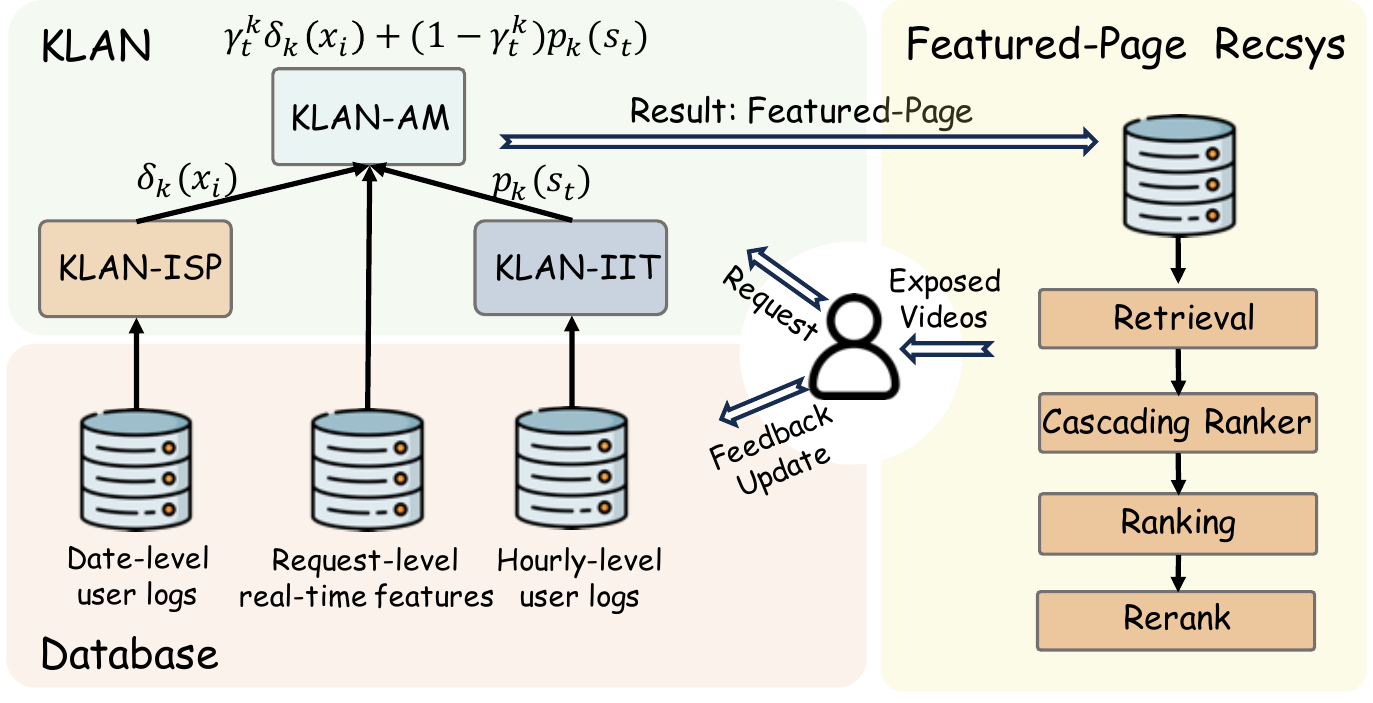}
    \vspace{-0.4cm}
\caption{Illustration of the online pipeline.}
	\label{fig:online_pipeline}
    \vspace{-0.5cm}
\end{figure}

\begin{figure*}[t]
   \setlength{\abovecaptionskip}{0.2cm}
   \setlength{\belowcaptionskip}{-0.0cm}
    \centering     \vspace{-0.25cm}
    \includegraphics[width=1.00\linewidth]{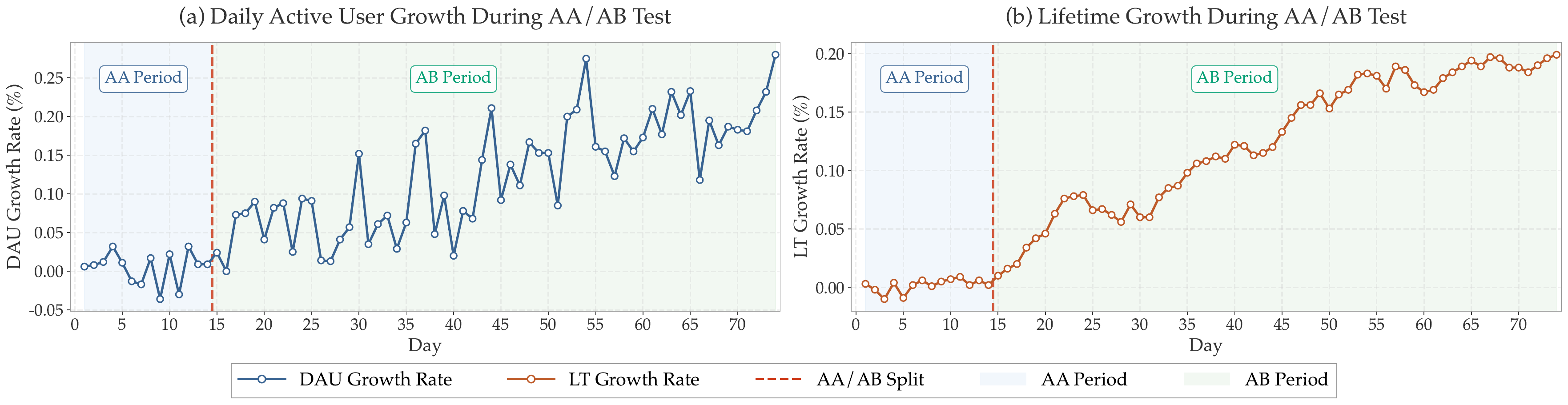}
    \vspace{-0.4cm}
    \caption{Online A/B test experimental results of DAU and LT.}
    \label{fig:online}
    \vspace{-0.25cm}
\end{figure*}

\subsubsection{\textbf{Hourly-level Data.}}

To model users’ intra-day dynamic interest transitions in the PLPM scenario, we construct an hourly-level dataset tailored for reinforcement learning (RL) tasks. This dataset is used to train the module KLAN-IIT, which captures temporal patterns in user behavior across app entry sessions. Each instance corresponds to a complete "app entry session" (defined as the sequence of interactions that starts with the display of a landing page when the user enters the app and ends when the user exits) and the required features of the corresponding next session. The dataset is organized in the standard RL format $(s_t, a_t, r(s_t,a_t), s_{t+1})$, where $a_t$ represents the assignment of a specific landing page upon app entry.
The state $s_t$ includes both real-time contextual features---such as current-day page-level consumption statistics, the number of live broadcasts by followed creators at the time of entry, and the number of prior app entries on the same day---and historical statistical features, including aggregated page consumption behavior over the past $n$ days. The reward $r(s_t,a_t)$ is defined as the app usage duration of the current session, serving as a proxy for user engagement. This dataset is constructed from hourly online logs collected in a real-world industrial environment, ensuring high temporal resolution and behavioral fidelity.

\subsubsection{\textbf{Real-time Streaming Data.}}
\label{sec:real_time_streaming}
In the AM module, we construct a real-time stream dataset derived from online user interaction logs to train the adaptive fusion model. This dataset is designed to facilitate the estimation of a fusion score that dynamically balances users’ static page preferences and dynamic interest transitions. The feature composition of the real-time dataset closely mirrors that of the hourly-level dataset, encompassing both real-time contextual signals and long-term behavioral statistics aggregated over the past $n$ days. Each instance corresponds to a user interaction event triggered by the delivery of a specific landing page. To characterize the user's immediate behavioral tendency, we define a binary supervision signal based on the ratio of total page switches to total app usage time during a session (from the user's entry to exit). If this ratio is less than a threshold $\mathcal{T}$, the user’s page interest is considered static and labeled as 1; otherwise, it is considered dynamic and labeled as 0. This label serves as a proxy to distinguish between user states dominated by long-term static preferences and those driven by transient interest shifts.



\begin{table}[t]
   \setlength{\abovecaptionskip}{0cm}
   \setlength{\belowcaptionskip}{-0.0cm}

\caption{The performance of Online A/B Testing at the platform. Boldface represents significance level $p$-value $<0.05$ of comparing KLAN with the online baseline. The values presented in the table represent the average observations of the last 7 days.}
    \centering
	\resizebox{0.88\linewidth}{!}{
    \begin{tabular}{c|c}
    \toprule
    {Dataset}&
    \multicolumn{1}{c}{Online Platform} \cr
    \cmidrule(lr){1-1}
    \cmidrule(lr){2-2}
    {Metrics} &\textbf{KLAN}\cr
    \midrule
    Daily Active Users &\textbf{+0.205\%}\cr
    Lifetime (LT) &\textbf{+0.192\%}\cr
    APP usage time &\textbf{+0.316\%}\cr
    Watch Time &\textbf{+0.443\%}\cr
    Video View &\textbf{+0.632\%}\cr
    Page Drop-off Ratio &\textbf{-28.158\%}\cr
    Featured Page Effective Entry Frequency &\textbf{+0.54\%}\cr
    Follow Page Effective Entry Frequency &\textbf{+2.56\%}\cr
    Explore Page Effective Entry Frequency &\textbf{+1.97\%}\cr
    Latency &-0.050\%\cr
    \bottomrule
    \end{tabular}}
    \label{online_ab111}
    \vspace{-6mm}
\end{table}

\subsection{Online A/B Test}

To further evaluate the online performance of our KLAN in a real-world production environment, we rebuilt the original personalized landing page service and directly deployed KLAN on the Kuaishou platform, which serves hundreds of millions of users. We then conducted an online experiment on approximately 5\% of the traffic for about 2.5 months, covering both the AA and AB testing phases.

\subsubsection{\textbf{System Description.}}
As shown in Figure \ref{fig:online_pipeline}, in the pipeline of the Kuaishou platform, when a user enters the app, a request is sent to the server, and then the personalized landing page service is called first. This service will process the historical interaction records and other information of the current user, and finally return the indicator of the specified page for this request, then the request has ended. After this, a new request will be generated based on the corresponding page indicator to trigger the corresponding services. Taking the featured-page recommendation service as an example, this request will send the corresponding user attributes and context features to the featured-page recommendation online service. Then trigger the online pipeline to filter and select videos from the candidate pool in a cascading form. Finally, after guiding the user to the corresponding page, the selected videos will be displayed to the user. Last but not least, when the user kills an app process in the background and re-enters the app, a new request will be generated again to invoke the personalized landing page service. 
The existing online baseline has evolved over years of iterative development and is composed of a large number of intricate heuristics—such as restoring the user's last exited page under specific conditions, selecting the most frequently visited page based on historical behavior, or choosing the page with the highest total engagement time. The intricate inter-dependencies among these strategies have made the system increasingly opaque, posing significant challenges for attribution, maintenance, and further optimization.

\begin{table}[t]
   \setlength{\abovecaptionskip}{0.1cm}
   \setlength{\belowcaptionskip}{-0.0cm}
\caption{The ablation result of Online A/B Testing at the platform. Boldface represents significance level $p$-value $<0.05$ of comparing KLAN variants with the online baseline.}
    \centering
	\resizebox{0.85\linewidth}{!}{
    \begin{tabular}{c|ccc}
    \toprule
    {Scenario}&
    \multicolumn{3}{c}{Online Platform} \cr
    \cmidrule(lr){1-1}
    \cmidrule(lr){2-4}
    {Metrics} &KLAN-IIT &KLAN-ISP &\textbf{KLAN}\cr
    \midrule
    APP usage time &\textbf{+0.067\%}&\textbf{+0.159\%}&\textbf{+0.316\%}\cr
    Watch Time &\textbf{+0.106\%}&\textbf{+0.318\%}&\textbf{+0.443\%}\cr
    Video View &+0.025\%&\textbf{+0.484\%}&\textbf{+0.632\%}\cr
    \bottomrule
    \end{tabular}}
    \label{online_ab222}
    \vspace{-3mm}
\end{table}

\subsubsection{\textbf{Scheme Overview.}}

Our model, KLAN, directly replaces the original personalized tab service to guide users to specific pages each time they open the app. As shown in Figure \ref{fig:online_pipeline}, KLAN-ISP and KLAN-IIT are trained using daily-level and hourly-level user data, respectively, while KLAN-AM is trained on real-time data streams.

During the inference phase, when a user's request triggers KLAN, the system retrieves the corresponding $\delta_k(\mathbf{x}_i)$ generated by KLAN-ISP from Redis using user ID.     At the same time, KLAN-IIT performs real-time inference to obtain the current $p_k(s_t)$ for the user. In addition, KLAN-AM is also invoked to estimate the value of $\gamma_t^k$. Based on these values and a predefined mapping $\sigma_k = \gamma_t^k \cdot \delta_k(\mathbf{x}_i) + (1 - \gamma_t^k) \cdot p_k(s_t)$, KLAN calculates the probability $\sigma_k$ for each candidate page and selects the one with the highest probability as the final result, which then triggers downstream components such as the recommendation system.

\subsection{Online Experimental Results}

\subsubsection{Online A/B test results}
We evaluate KLAN using common consumption metrics (APP usage time, watch time, and daily active users), commercial metrics (video view), PLPM-specific metrics (page drop-off ratio), platform constrain metrics (effective page entry frequency and latency), and retention metrics (LT) to comprehensively assess its online performance.
Among these metrics, \textbf{LT} serves as a critical indicator of user retention and experience quality \cite{meng2024coarsetofinedynamicupliftmodeling}, which can be formulated as:
\begin{equation}
    \text{LT} = \frac{\sum_{i = \max(T - 6, T_0)}^T \text{DAU}_i}{\text{Total number of active users from } T_0 \text{ to } T}
    \label{equ:LT7}
\end{equation}
Figure \ref{fig:online} illustrates the temporal evolution of DAU and LT growth rates throughout the 74-day experimental period, comprising a 14-day AA test followed by a 60-day AB test.
During the AB test, we can find that KLAN achieves significant improvements in both metrics.
Table \ref{online_ab111} reports statistically significant improvements across all metrics compared to the online baseline.
Notably, APP usage time, watch time and video view counts show substantial growth, while page drop-off ratio decreases significantly, indicating enhanced user satisfaction with assigned landing pages. 
Regarding platform constraint metrics, the effective page entry frequency metrics for all three pages show positive improvements.
Latency results indicate our model maintains real-time efficiency with negligible overhead.
Currently, KLAN is fully deployed on Kuaishou, serving hundreds of millions of daily active users every day.

\begin{figure}[t]
   \setlength{\abovecaptionskip}{0.3cm}
   \setlength{\belowcaptionskip}{-0.0cm}
\includegraphics[scale=0.3]{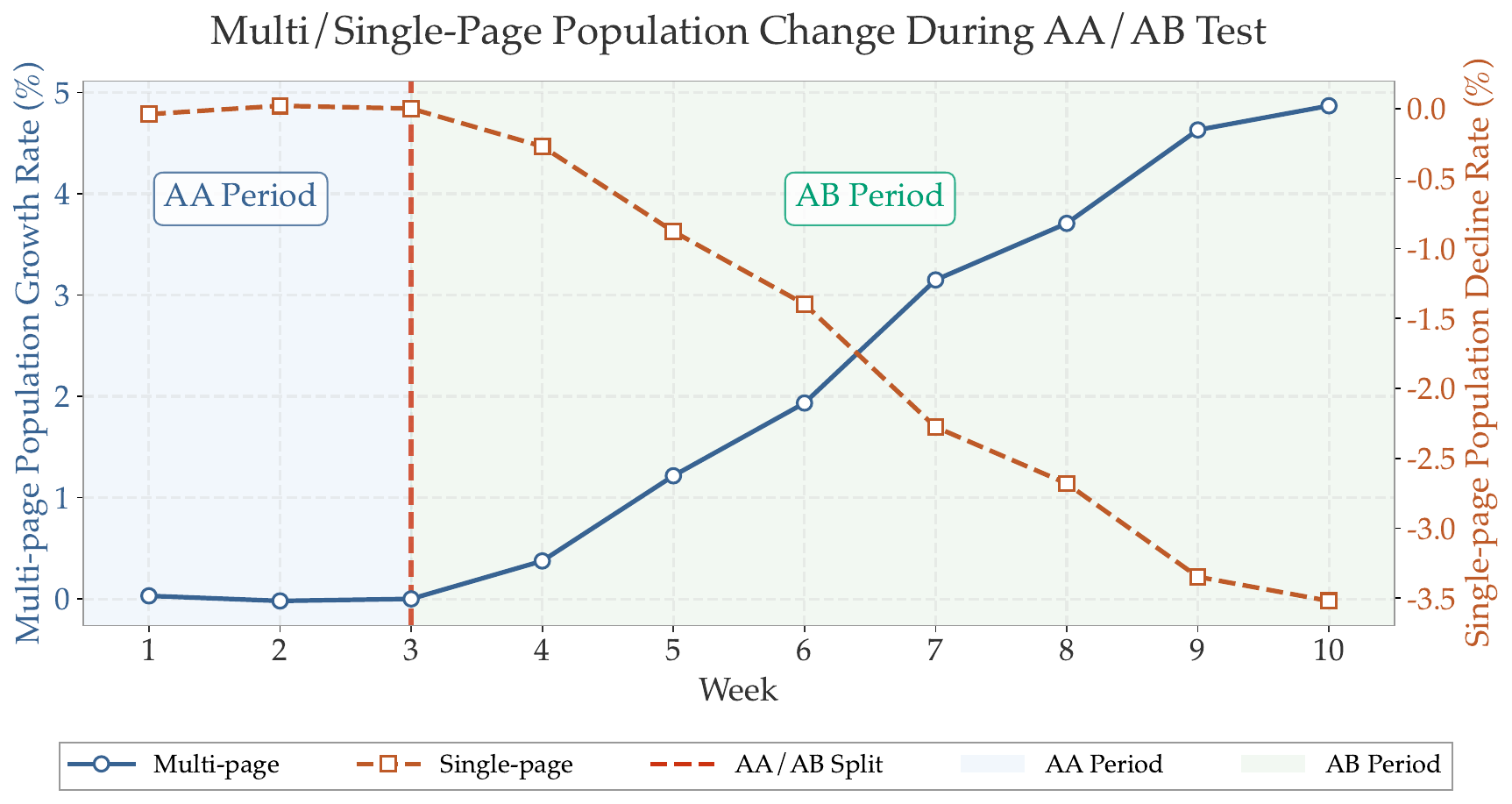}
\vspace{-0.5cm}
\caption{User transfer between single-page and multi-page users during AB period.}
\label{fig:indepth}
\vspace{-0.45cm}
\end{figure}

\subsubsection{Ablation Test of KLAN}
In this section, we perform ablation studies to evaluate the contribution of different components in KLAN, as shown in Table \ref{online_ab222}. Specifically, we compare KLAN with two variants: \textbf{(1) KLAN-IIT}, which relies solely on the IIT module using intra-day dynamic interest scores $\{p_{k}\}_{k=1}^K$ for page navigation decisions, and \textbf{(2) KLAN-ISP} which uses only the ISP module based on inter-day static preference scores $\{\delta_{k}\}_{k=1}^K$.

Both individual modules outperform the baseline, highlighting the value of modeling multi-temporal (inter-day and intra-day) user signals. Notably, KLAN-ISP captures users' inter-day static preferences, providing a strong foundation for navigation, while KLAN-IIT models intra-day dynamics, yielding consistent gains. Most importantly, KLAN significantly surpasses both variants, validating our hierarchical design: ISP offers stable preferences, IIT captures short-term transitions, and their adaptive fusion via the AM module enables synergistic performance beyond individual contributions. These consistent improvements confirm KLAN’s effectiveness in balancing long-term and short-term user interests for optimal page navigation.

\subsubsection{In-depth Analysis}
To evaluate KLAN’s impact, we analyze its effect on user navigation patterns—specifically, the transition from single-page to multi-page usage (Figure \ref{fig:indepth}). A/B test results reveal a clear shift: the proportion of single-page users decreases while multi-page users increase, suggesting that KLAN effectively encourages broader engagement through adaptive page assignment at app entry. Crucially, as shown in Figure \ref{fig:DA}-a, multi-page users exhibit significantly higher DAU and LT compared to single-page users. This shift in user composition thus explains the observed gains in both short-term activity and long-term retention, highlighting KLAN’s role in driving sustained platform growth.

\section{Conclusion and Future Work}
In this paper, we formally introduce Personalized Landing Page Modeling (PLPM) as a critical yet underexplored task in recommender systems.
To address it, we propose KLAN, a hierarchical framework that integrates inter-day static page preferences with intra-day dynamic interests through adaptive modulation in real-time.
Extensive online experiments on Kuaishou demonstrate KLAN's effectiveness and scalability. Moreover, our model is now deployed online with full users at Kuaishou platform, serving for hundreds of millions of users every day.
Finally, we will also release a well-established benchmark to encourage future innovations in the PLPM scenario.
\normalem
\bibliographystyle{ACM-Reference-Format}
\balance
\bibliography{reference}
\nocite{*}

\newpage

\appendix

\section{APPENDIX}

\begin{figure}[H]
	\centering
	\subfigure[Featured Page.]{\includegraphics[height=.6\linewidth]{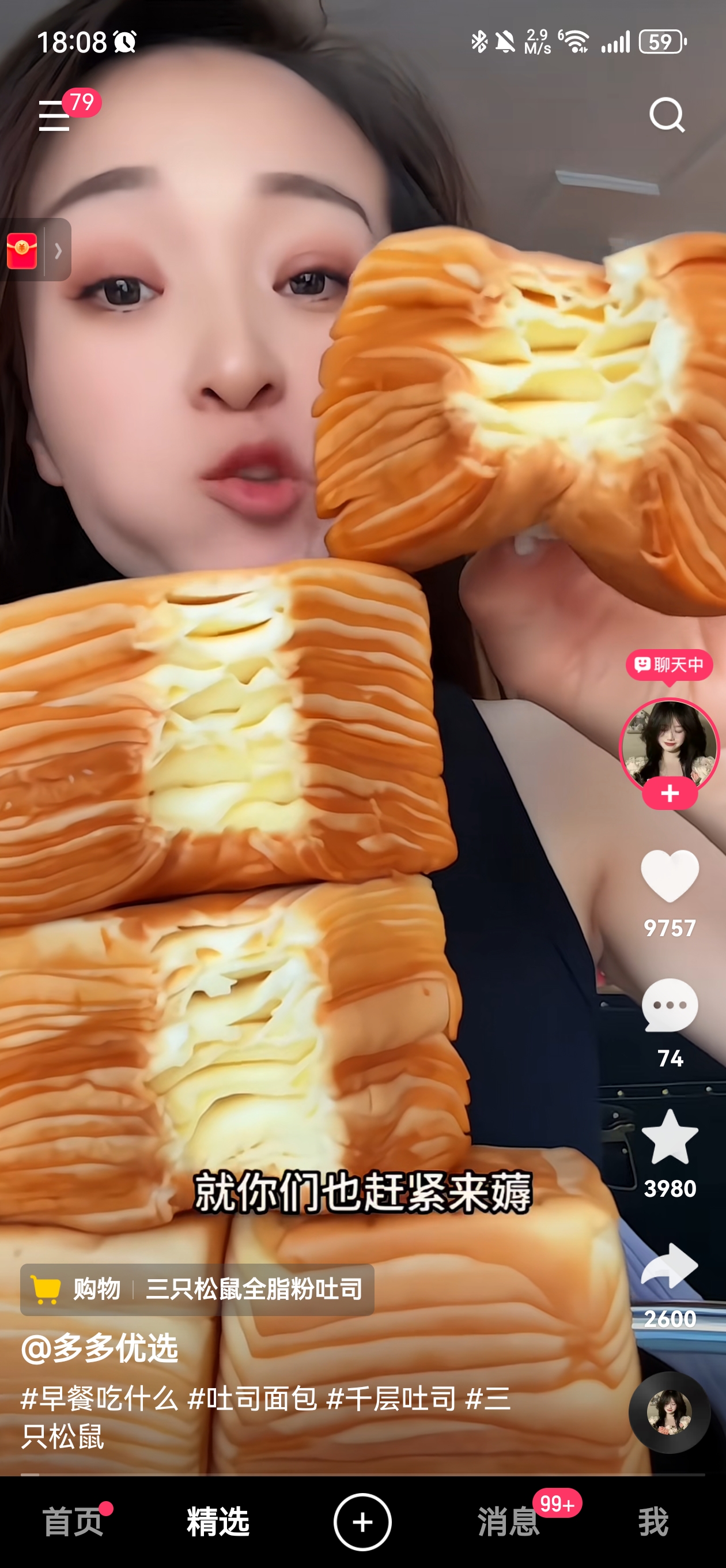}}\hspace{0.3cm} 
	\subfigure[Store Page.]{\includegraphics[height=.6\linewidth]{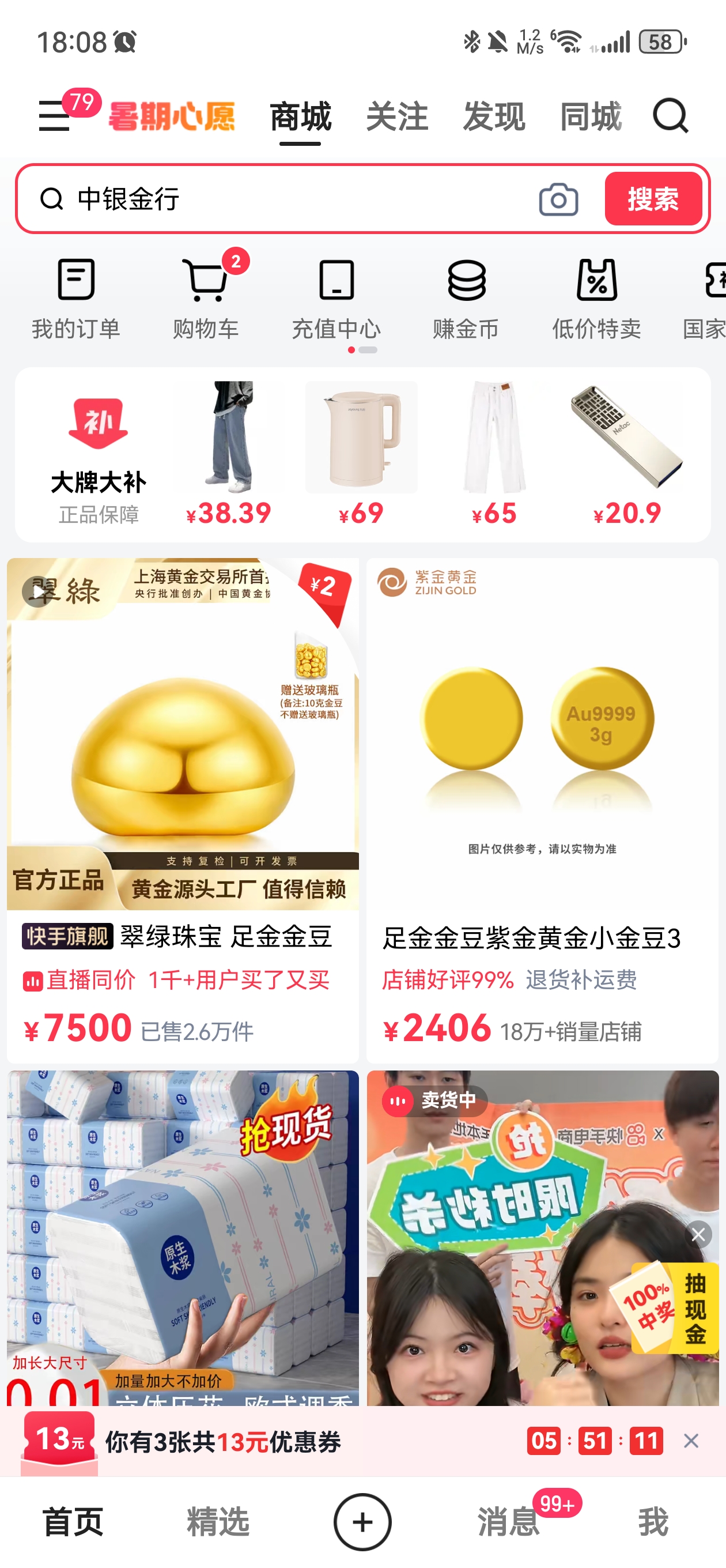}}\hspace{0.3cm} 
    \subfigure[Following Page.]{\includegraphics[height=.6\linewidth]{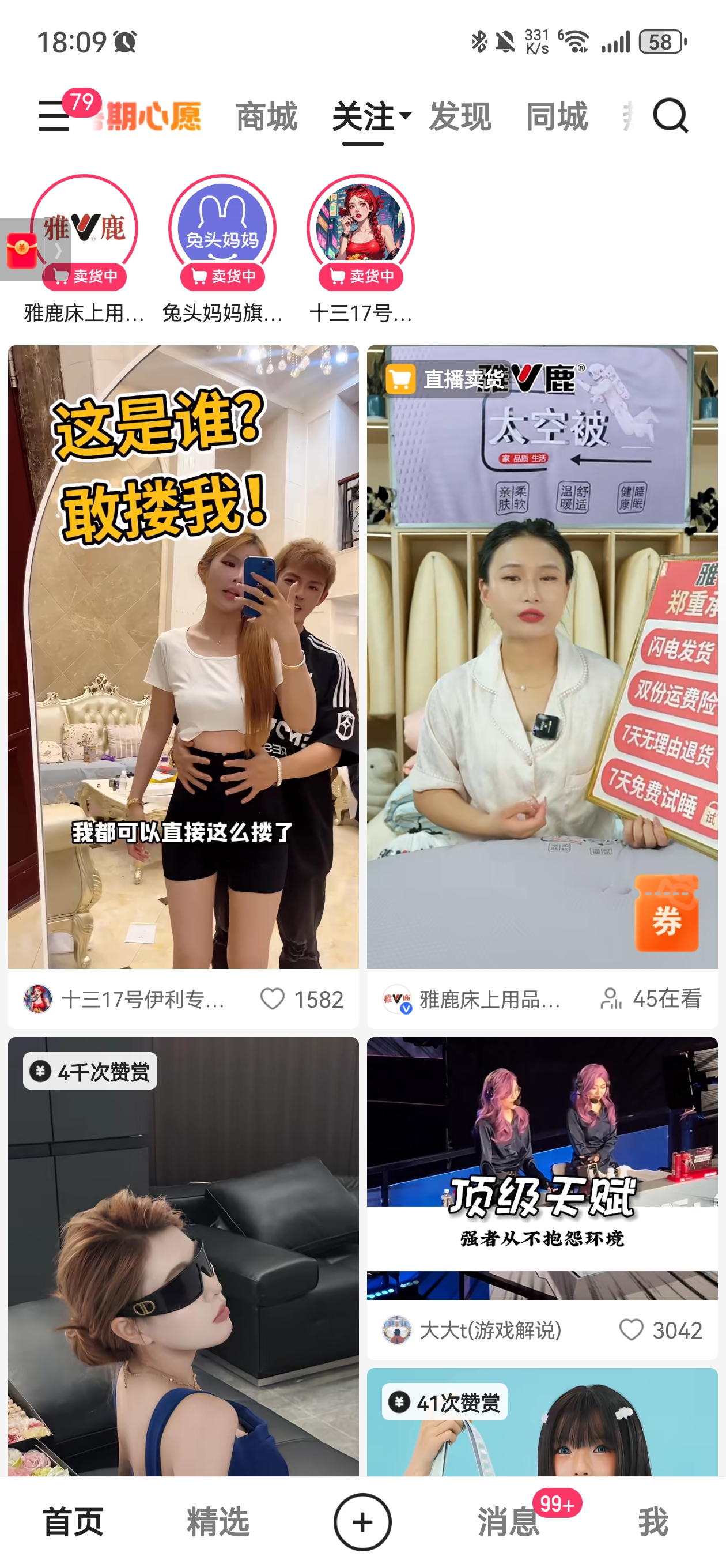}}
	\caption{Cases of different pages on Kuaishou APP.}
	\label{fig:platform_description}
\end{figure}

\subsection{Platform Description}
\label{sec:page_heterogeneity}
To satisfy different users interests, Kuaishou APP offers multiple functional and channel pages with distinct characteristics, leading to significant page heterogeneity.
This heterogeneity spans multi-dimensions (layout, content, and interaction patterns) and results in vastly different user response distributions across pages. 
Figure \ref{fig:platform_description} illustrates three representative page types that demonstrate this principle.

\textbf{Featured Page} employs a full-screen, single-column layout optimized for immersive content consumption. 
Users can scroll vertically to seamlessly transition between videos, while strategically placed interaction buttons (like, comment, follow) encourage real-time engagement. 
In contrast, \textbf{Store Page} adopts a double-column layout, similar to traditional e-commerce platforms, prioritizing commercial efficiency over entertainment. Clear product listings with prices enable users to quickly browse and compare multiple items. The interface emphasizes transactional actions—clicking products, viewing details, adding to cart—reflecting its commerce-oriented objectives.
\textbf{Following Page} strikes a balance between the two approaches, utilizing a double-column format that presents content from following creators. 
This layout optimizes for both content discovery and social connection, allowing users to efficiently browse updates and live streaming information from familiar creators while maintaining the personal touch through likes and comments. 
Such page diversity, while essential for user experience, introduces substantial complexity for recommendation systems. The divergent user behaviors and response patterns across these pages challenge traditional joint-learning approaches, necessitating more sophisticated modeling strategies to effectively capture page-specific preferences.

\subsection{Feature Embedding and Selection}
\label{sec:feature_selection}
For each instance $z_i = (\mathbf{x}_i, \mathbf{t}_i, y_i)$, we transform the original features into dense embeddings through embedding lookup operations:
\begin{equation}
\mathbf{e}_{xi} = \mathbf{E}_x^{T} \cdot \mathbf{x}_i, \quad 
\mathbf{e}_{ti} = \mathbf{E}_t^{T} \cdot \mathbf{t}_i,
\end{equation}
where $\mathbf{E}_x \in \mathbb{R}^{f_x \times d}$ and 
$\mathbf{E}_t \in \mathbb{R}^{f_t \times d}$ are the created embedding tables, 
$f_x$ and $f_t$ are the numbers of non-treatment features and treatment features,  and $d$ is the embedding size.

To perform treatment-specific feature selection, we employ a target attention mechanism that extracts relevant features from the non-treatment embedding $\mathbf{e}_{xi} $ guided by the treatment embedding $\mathbf{e}_{ti}$:
\[
\left\{
\begin{aligned}
\mathbf{w}_i^k &= Softmax\left(\mathbf{W}^k \cdot \mathbf{e}_{ti} + \mathbf{b}^k\right) \\
\mathbf{e}_{xi}^k &= \sum_{j=1}^{f_x} \left( \mathbf{w}_i^k(j) \cdot \mathbf{e}_{xi,j} \right)
\end{aligned}
\right.
\]
where $\mathbf{e}_{xi}^k$ denotes the extracted non-treatment embedding for the $k$-th (corresponding to $t_i$) treatment. 
$\mathbf{W}^k \in \mathbb{R}^{f_x \times d}$ and $\mathbf{b}^k \in \mathbb{R}^{f_x \times 1}$ are feature transformation matrix and bias matrix. 
And $\mathbf{w}_i^k \in \mathbb{R}^{f_x \times 1}$ is the attention vector which is used as selector to calculate the weighted sum of all non-treatment embeddings. 
$\mathbf{w}_i^k(j)$ denotes the $j$-th element of vector $\mathbf{w}_i^k$, and $\mathbf{e}_{xi,j}$ is the $j$-th non-treatment feature. 
It is worth noting that in the case of $k = 0$, the tree structure we designed follows a multi-branch joint optimization approach. 
Therefore, the parameters $\mathbf{W}^k$ and $\mathbf{b}^k$ under the corresponding branch are shared with $k$ under specific values.

\subsection{KLAN-ISP Theoretical Analysis}
\label{sec:theoratical}
In this section, we first provide a theoretical analysis to demonstrate the coupled gradient issue inherent in existing uplift models based on shared input representations.
Subsequently, we present a contrasting proof for our proposed KLAN-ISP, demonstrating how its architecture achieves gradient decoupling. 
This analysis theoretically validates the superiority of our approach in addressing gradient conflicts when handling multi-treatment heterogeneous datasets in the PLPM scenario.

\begin{figure}[H]
	\centering
	\includegraphics[width=0.3\textwidth]{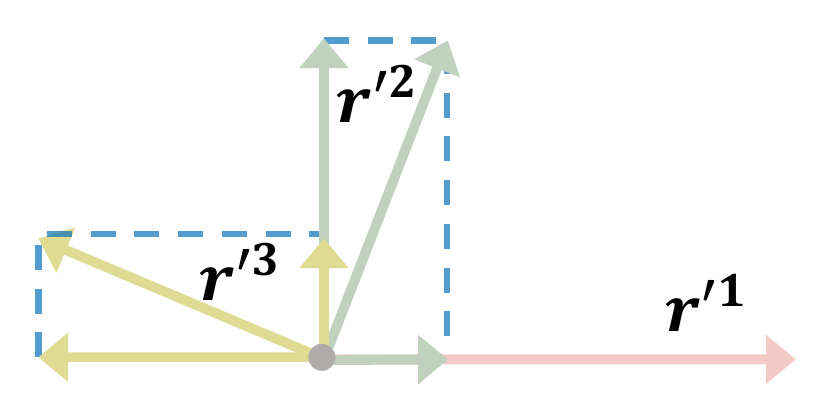}
    \Description{An example of the gradient conflict.}
	\caption{An example of the gradient conflict ($K=3$).}
	\label{fig:conflict}
\end{figure}

\textbf{Proof of The Coupled Gradient Issue in Existing Uplift Methods:}
For simplicity, we assume that the learned non-treatment and treatment representations in existing uplift methods can be expressed as
\begin{equation}
    \textbf{e}_x = g_x(\textbf{x}), \textbf{e}_t = g_t(\textbf{t})
\end{equation}
where $g_x(\cdot)$ and $g_t(\cdot)$ denote the representation learning functions for non-treatment features $\mathbf{x}$ and treatment features $\mathbf{t}$, respectively.

These representations are subsequently concatenated to construct a unified feature representation:
\begin{equation}
    \textbf{f} = \text{concat}(\textbf{e}_x,\textbf{e}_t)
\end{equation}

In the existing uplift methods based on traditional multi-task learning paradigm, this shared representation $\mathbf{f}$ serves as \textbf{\textsl{same
input}} to $K$ task-specific predictors $\{h_k(\cdot)\}_{k=1}^K$, each estimating the outcome for a specific treatment. The aggregate loss function is defined as:
\begin{equation}
\begin{aligned}
\mathcal{L}
&=\sum_{k=1}^{K}L(\hat{y}^{k}-{y}^{k})\\
&=\sum_{k=1}^{K}L(h_{k}(\mathbf{f})-{y}^{k}),
\end{aligned}
\end{equation}
where $\hat{y}^k = h_k(\mathbf{f})$ represents the prediction for the $k$-th treatment, and $y^k$ denotes the corresponding ground truth.

To demonstrate the phenomenon of gradient coupling, we examine the gradient of $\mathcal{L}$ with respect to the shared representation $\mathbf{f}$:

\begin{equation}
\begin{aligned}
{\partial{\mathcal{L}}\over{\partial{\mathbf{f}}}}
&=
{\partial\over{\partial{\mathbf{f}}}}\left(\sum_{k=1}^{K}L(h_{k}(\mathbf{f})-{y}^{k})\right)\\
&=
\sum_{k=1}^{K}{\partial{L(h_{k}(\mathbf{f})-{y}^{k})}
\over{\partial{\mathbf{f}}}}\\
&=
\sum_{k=1}^{K}{\partial{h_{k}(\mathbf{f})}\over{\partial{\mathbf{f}}}} \cdot {L^{'}(h_{k}(\mathbf{f})-{y}^{k})}\\
&=
\sum_{k=1}^{K}{a^{k}{\mathbf{r}^{k}}}\\
&=
\sum_{k=1}^{K}{\mathbf{r}^{' k}},
\end{aligned}
\end{equation}
where $a^{k} = L^{'}(h_{k}(\mathbf{f})-{y}^{k})$ represents a scalar. We also define $\mathbf{r}^{k} = {\partial{h_{k}(\mathbf{f})}\over{\partial{\mathbf{f}}}}$ as a vector that indicates the update direction preferred by task $k$. Consequently, $\mathbf{r}^{' k}$ represents the complete gradient vector contributed by task $k$, encapsulating both magnitude and direction.

This formulation reveals that the parameter update for the shared representation $\mathbf{f}$ is determined by the superposition of all task-specific gradients $\{\mathbf{r}'^k\}_{k=1}^K$. When these gradient vectors are not aligned, conflicting optimization objectives emerge. 

To quantify this misalignment, consider an orthogonal decomposition of gradient vectors $\{\mathbf{r}'^k\}_{k=2}^K$ with respect to a reference gradient $\mathbf{r}'^1$ (Figure \ref{fig:conflict}). The presence of orthogonal components indicates that different tasks impose conflicting constraints on the shared representation, leading to suboptimal convergence. This gradient coupling phenomenon fundamentally limits the effectiveness of naive parameter sharing in multi-task learning for treatment effect estimation, as the shared representation must simultaneously optimize potentially contradictory objectives across all treatment tasks.

\textbf{Proof of Decoupled Gradient of KLAN-ISP}
We now prove that KLAN-ISP achieves gradient decoupling by confining each sample's gradient flow to its corresponding treatment branch.
Consider a training sample $(\mathbf{x}_i, \mathbf{t}_i, y_i)$ where treatment $\mathbf{t}_i \in {1, 2, \ldots, K}$ was applied. Unlike traditional uplift methods, KLAN-ISP employs separate inputs $\mathbf{f}_i^k$ for the $k$-th treatment-specific branch, where $\mathbf{f}_i^k$ is derived from both $\mathbf{x}_i$ and $\mathbf{t}_i$ (see Appendix \ref{sec:feature_selection}).

\textit{Loss Function}
For a sample $(\mathbf{x}_i, \mathbf{t}_i, y_i)$ that received treatment $\mathbf{t}_i$, 
the loss function is:
\begin{equation}
\mathcal{L}_i = \left( y_i - \hat{y}_i^{\mathbf{t}_i} \right)^2
\end{equation}
where the prediction $\hat{y}_i^{\mathbf{t}_i}$ is the output of the $\mathbf{t}_i$-th branch. Crucially, $\hat{y}_i^{\mathbf{t}_i}$ is a function of the internal components of its own branch, which are all derived from the branch's specific input representation $\mathbf{f}_i^{\mathbf{t}_i}$.

\textit{Gradient Derivation:}
To prove gradient decoupling, we compute the partial derivative of the loss $\mathcal{L}_i$ with respect to the input representation of an arbitrary branch $k$, denoted as $\mathbf{f}_i^k$. We will demonstrate that this derivative is non-zero only when $k = \mathbf{t}_i$ and is strictly zero otherwise.

\textit{Case 1: Gradient with respect to the active branch's input ($k=\mathbf{t}_i$).}
First, we compute the derivative of the loss with respect to the input representation of the branch that corresponds to the applied treatment, $\mathbf{f}_i^{\mathbf{t}_i}$. Applying the chain rule, we have:
\begin{equation}
\begin{aligned}
\frac{\partial \mathcal{L}_i}{\partial \mathbf{f}_i^{\mathbf{t}_i}} &= \frac{\partial \mathcal{L}_i}{\partial \hat{y}_i^{\mathbf{t}_i}} \cdot \frac{\partial \hat{y}_i^{\mathbf{t}_i}}{\partial \mathbf{f}_i^{\mathbf{t}_i}} \\
&= -2(y_i - \hat{y}_i^{\mathbf{t}_i}) \cdot \frac{\partial h^{\mathbf{t}_i}(\mathbf{z}_i^{\mathbf{t}_i})}{\partial \mathbf{f}_i^{\mathbf{t}_i}}
\end{aligned}
\end{equation}
Since the prediction error $(y_i - \hat{y}_i^{\mathbf{t}_i})$ and the gradient of the $\mathbf{t}_i$-th branch's network are generally non-zero during training, the overall gradient is non-zero:
\begin{equation}
    \frac{\partial \mathcal{L}_i}{\partial \mathbf{f}_i^{\mathbf{t}_i}} \neq 0
\end{equation}
This confirms that the gradient updates occur through the active branch.

\textit{Case 2: Gradient with respect to an inactive branch's input ($k \neq \mathbf{t}_i$).}
We now examine the gradient behavior for inactive branches—those that do not correspond to the applied treatment. For any branch $k$ where $k \neq \mathbf{t}_i$, we compute the derivative of the loss with respect to its input representation $\mathbf{f}_i^k$:
\begin{equation}
\frac{\partial \mathcal{L}_i}{\partial \mathbf{f}_i^k} = \frac{\partial \left( y_i - \hat{y}_i^{k} \right)^2}{\partial \mathbf{f}_i^k}
\end{equation}
During training, the model masks the outputs of all branches except the one corresponding to the applied treatment and $y_i$ is a constant (the ground truth). Therefore:
\begin{equation}
\frac{\partial \mathcal{L}_i}{\partial \mathbf{f}_i^k} =0, \quad \text{for } k \neq \mathbf{t}_i
\end{equation}

The derivation formally proves that for any sample under a specific treatment $\mathbf{t}_i \neq 0$, the gradient of the loss function exists only for the corresponding $\mathbf{t}_i$-th branch, while the gradients for all other branches are zero. This demonstrates a \textbf{strong and complete gradient decoupling}. The learning of each treatment-specific branch is isolated, which structurally prevents the gradient conflicts arise in existing uplift methods.

\begin{figure*}
    \centering     
    \includegraphics[width=1.00\linewidth]{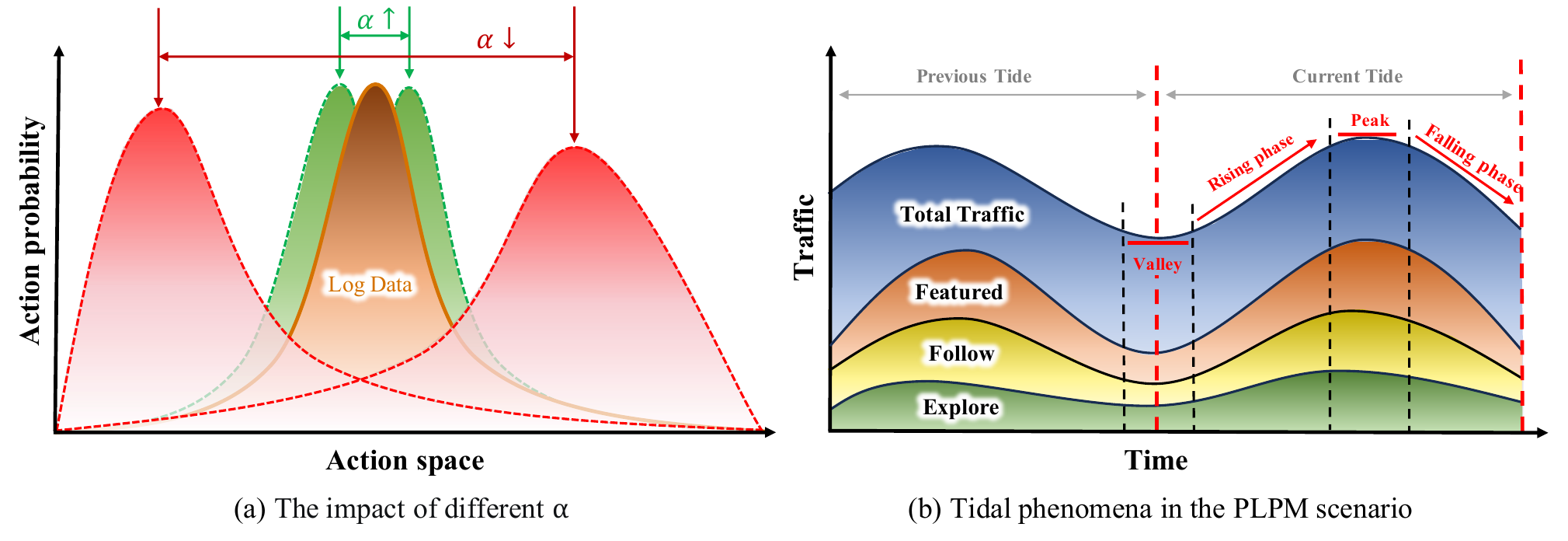}
    \caption{Illustration of different $\alpha$ for CQL and tidal phenomenon.}
    \label{fig:rl_illustration}
\end{figure*}

\subsection{KLAN-IIT Design Details}
\label{sec:TIDAL}
In CQL, the conservative coefficient $\alpha$ determines the trade-off between policy exploration and exploitation, as illustrated in Figure \ref{fig:rl_illustration}-a.
To elaborate, an excessively large conservative coefficient leads to an overly pessimistic policy (the green region) that is tightly constrained to the log data, ensuring safety but sacrificing the policy exploration.
Conversely, a small coefficient permits a more aggressive policy (the red region) that encourages policy exploration, but at the risk of instability.

However, the efficacy of any static $\alpha$ is fundamentally challenged in our proposed PLPM scenario. 
We find that the log data does not originate from steady-state policy distributions but exhibits a distinct intra-day "Tidal Phenomenon". 
As illustrated in Figure \ref{fig:rl_illustration}-b, both total traffic and page-specific traffic exhibit periodic fluctuations within daily cycles, with the confidence level of samples varying accordingly.
During peak phases, abundant and consistent user interactions yield training samples with high statistical reliability; whereas in valley phases, sparse and noisy traffic results in data with significantly lower confidence level.
This periodic variability presents a fundamental challenge for traditional Conservative Q-Learning (CQL), as its static, uniform conservative coefficient $\alpha$ cannot adapt to the fluctuating confidence level inherent in intra-day data patterns.

To mitigate the aforementioned issues, we introduce the Dynamic Conservative Coefficient $\alpha_t$, aiming at effectively balancing the exploration-exploitation trade-off in intra-day RL modeling.
To be specific, during periods of high data stability and reliability (i.e., peak phases), we adaptively decrease $\alpha_t$ to encourage more aggressive policy optimization from high-quality data. Conversely, when the data distribution is sparse or exhibits significant shift (i.e., valley phases), we increase $\alpha_t$ to enhance conservatism, thereby safeguarding the policy against performance degradation arising from data noise or distribution shift.

To operationalize the dynamic conservative coefficient $\alpha_t$, we leverage several key features correlated with the "tidal phenomenon" in the PLPM scenario. For example, app entry time, daily app entry frequency, etc.

Take app entry time as an example, we have the following analysis:
\begin{itemize}
    \item High activity periods (12-14h, 19-22h): Large user volume with dense interaction data (e.g., dwell time). User preference estimation is more accurate with lower uncertainty, requiring weaker regularization (smaller coefficient).
    \item Low activity periods (3-6h): Sparse user volume and limited data. Preference estimation contains higher noise and uncertainty, requiring stronger regularization (larger coefficient).
\end{itemize}
We first compute the average interaction rate $V_t$ for each hour based on historical data:
\begin{equation}
V_t = \frac{24*\text{The total usage time of the $t$-th hour}}{\text{The total usage time of the whole day}}
\end{equation}
To adaptively adjust the regularization strength across different hours, we define the hour-based regularization weight $h_t$ as:
\begin{equation}
h_t = \alpha \cdot e^{-\beta (V_t - 1)}
\end{equation}
where $V_t$ denotes the relative activity level at hour $t$, and $\alpha$, $\beta$ are hyperparameters. This formulation ensures that hours with lower user activity (i.e., $V_t < 1$) receive stronger regularization, while more active hours (i.e., $V_t > 1$) are regularized less aggressively.


\textbf{Feature Combination:}
The final dynamic coefficient combines both features as followed:
\begin{equation}
    \alpha_t = \alpha \times h_t,
\end{equation}
where $\alpha$ is the base coefficient. We constrain $\alpha_t$ within bounds of $[0.1\alpha, 5\alpha]$ to ensure stability.

\subsection{KLAN-AM Details}
\label{sec:AM_training}
For the selection of features and labels, as mentioned in Section \ref{sec:def_AM}, we use contextual features $\mathbf{c}_t$ and the users’ long-term behavioral statistics $\mathbf{v}_t$ as features. And a binary supervision signal $\boldsymbol{\gamma}_t^{k_t}$, is based on the ratio of total page switches to total app usage time during a session (from the user's entry to exit). If this ratio is less than a threshold $\mathcal{T}$, the user’s page interest is considered static and labeled as 1; otherwise, it is considered dynamic and labeled as 0.

To learn the context-aware weights, we first encode the input features into embeddings. 
We then employ a multi-task learning framework based on MMOE~\cite{ma2018modeling}, where each expert network specializes in predicting the weight for a specific landing page.
The output of tower $k$ represents the users’ immediate behavioral tendency, indicating the relative importance of static preferences for landing page $k$. During training, we optimize the model using Cross Entropy loss. For each training instance $ (\mathbf{c}_t,\mathbf{v}_t, \mathbf{k}_t,\boldsymbol{\gamma}_t^{k_t})$, we only compute the loss for the tower corresponding to the assigned landing page $k_t$, while the outputs from other towers are masked.


\subsection{Computational Complexity Discussion}

\subsubsection{KLAN-ISP}

Let $N_1$ denote the batch size, $K$ the number of treatments, $M$ the number of experts per branch, $d$ the embedding dimension, $h$ the latent representation dimension, and $f_x$ the number of non-treatment features.
For each training instance, KLAN-ISP performs: (1) feature embedding and treatment-specific feature selection with complexity $O(K\cdot f_x \cdot d)$; (2) forward propagation through $K$ branches, where each branch generates $M$
 expert representation of dimension $h$ from $d$ dimensional inputs, and performs weighted aggregation, resulting in $O(K \cdot M \cdot d \cdot h)$ complexity; (3) tower compute complexity $C_{tower}$ (4)  loss computation including MSE and KL divergence constraints on $h$ dimensional representations across branches with $O(K \cdot h)$. 
 Therefore the overall training complexity of a batch is $O(N_1 \cdot K \cdot (f_x\cdot d + M \cdot d \cdot h + C_{tower}))$.
 Regarding inference phase, for each instance, the model performs $K$ forward passes to obtain all treatment effects, followed by averaging operations to compute the final inter-day static page preference scores. The inference complexity is $O(K \cdot (f_x \cdot  d+M \cdot d \cdot h + C_{tower}))$.
 
 Notably, KLAN-ISP operates on daily-level data with updates occurring only once per day.
 After training, the model performs inference on all instances, and the resulting inter-day static page preference scores are stored for subsequent direct usage.

\subsubsection{KLAN-IIT}
Let $N_2$ denote the batch size of KLAN-IIT,  $K$ the number of landing pages, $d_s$ the dimension of state representation, and $d_h$ the dimension of hidden layer in the Q-networks.
During each training step, KLAN-IIT processes a batch of $N_2$ instances. For each instance, KLAN-IIT performs: (1) forward passes through both the main Q-network and target Q-network, each with complexity $O(d_s \cdot d_h + d_h\cdot K)$; (2) TD loss calculation, which requires evaluating all $K$ actions to find $\max_{a'} Q(s_{t+1}, a')$ with complexity $O(K)$ and loss calculation $O(1)$; (3) conservative regularization, which computes the LogSumExp over all actions with complexity $O(K)$.
The overall train complexity of per batch is $O(N_2 \cdot (d_s \cdot d_h + d_h\cdot K))$.
During inference, for each instance, the model performs forward proceed to access the $K$ Q-value, followed by a softmax normalization. The overall complexity is $ O(d_s \cdot d_h + d_h \cdot K)$.

\subsubsection{KLAN-AM}
Let $E$ denote the number of experts in the MMOE architecture, $d_c$ the dimension of contextual features, $d_v$ the dimension of page features and $d_{moe}$ the hidden dimension of each expert network.
For each inference instance, KLAN-AM performs: (1) context-aware weight estimation through MoE, which involves $E$ experts proceeds with complexity $O(E\cdot ((d_c + d_v) \cdot d_{moe}+ d_{moe} \cdot K))$ and weighted aggregation with $O(E \cdot K)$; (2) adaptive score fusion of inter-day static page preference scores and intra-day dynamic interest scores using context-aware weights, resulting in $O(K)$ complexity; (3) page assignment through maximum selection with complexity $O(K)$.
Therefore, the overall inference complexity is $O(E\cdot ((d_c + d_v) \cdot d_{moe}+ d_{moe} \cdot K))$.

\newpage
\onecolumn
\subsection{Notations and Corresponding Descriptions}

\begin{table}[h]
\centering
\caption{RCT Dataset and Basic Notations}
{\renewcommand{\arraystretch}{1.2}
\begin{tabular}{ll}
\toprule
\textbf{Symbol} & \textbf{Description} \\
\midrule
$\mathcal{Z}$ & Daily-level dataset collected from Randomized Controlled Trial (RCT) \\
$N$ & Number of all instances in $\mathcal{Z}$ \\
$z_i = (\mathbf{x}_i, \mathbf{t}_i, y_i)$ & The $i$-th instance comprising user feature, treatment, and user response \\
$\mathbf{x}_i$ & User feature of the $i$-th instance \\
$\mathbf{t}_i \in [1, K]$ & Specific treatment of the $i$-th instance \\
$y_i$ & Observed user response of the $i$-th instance \\
$K$ & Number of available landing pages/treatments \\
\bottomrule
\end{tabular}
}
\end{table}

\begin{table}[h]
\centering
\caption{Causal Inference Notations}
{
\renewcommand{\arraystretch}{1.2}
\begin{tabular}{ll}
\toprule
\textbf{Symbol} & \textbf{Description} \\
\midrule
$y_i^{\mathbf{t}_i}$ & Response when user in $i$-th instance receives treatment $\mathbf{t}_i$ \\
$y_i^{\mathbf{t}_0}$ & Response when user in $i$-th instance receives no treatment \\
$\tau^k(\mathbf{x}_i)$ & Expected individual treatment effect for treatment $k$ \\
$\mathbb{E}[y_i^k \mid \mathbf{t}_i = k, \mathbf{x}_i]$ & Expected response under treatment $k$\\
$\mathbb{E}[y_i^0 \mid \mathbf{t}_i, \mathbf{x}_i]$ & Expected response under no treatment \\
\bottomrule
\end{tabular}
}
\end{table}

\begin{table}[h]
\centering
\caption{MDP Framework Notations}
{
\renewcommand{\arraystretch}{1.2}
\begin{tabular}{ll}
\toprule
\textbf{Symbol} & \textbf{Description} \\
\midrule
$\mathcal{S}$ & Continuous representation space of user state \\
$\mathcal{A}$ & Action space comprising all available landing page assignments \\
$\mathcal{P}: \mathcal{S} \times \mathcal{A} \rightarrow \mathcal{S}$ & State transition function \\
$r(s_t, a_t)$ & Immediate reward that captures the user feedback for the landing-page assignment $a_t \in \mathcal{A}$ on user state $s_t \in \mathcal{S}$. \\
$\gamma \in [0,1]$ & Discount factor balancing immediate and future rewards \\
$s_t$ & The representation of user's real-time and daily features upon app entry at step $t$ \\
$a_t$ & The representation of the assignment of specific landing page upon app entry at step $t$ \\
$\pi: \mathcal{S} \rightarrow \mathcal{A}$ & Policy function \\
$\pi^*$ & Optimal policy maximizing expected discounted return \\
\bottomrule
\end{tabular}
}
\end{table}


\begin{table}[h]
\centering
\caption{KLAN-ISP Specific Notations}
{
\renewcommand{\arraystretch}{1.2}
\begin{tabular}{ll}
\toprule
\textbf{Symbol} & \textbf{Description} \\
\midrule
$e_{xi}$, $e_{ti}$ & Embeddings of non-treatment and treatment features \\
$\mathbf{e}_{xi}^k$ & Treatment-specific selected features from target attention \\
$\text{FS}_k(\cdot)$ & $k$-th treatment-specific feature selection module \\
$\mathbf{f}_i^k$ & Input feature vector for $k$-th branch \\
$M$ & Number of expert \\
$\mathbf{d_{im}^k}$ & $m$-th expert representation of $i$-th instance in $k$-th branch \\
$\mathbf{g}_i^k \in \mathbb{R}^{M \times 1}$ & Gating weights with $\mathbf{g}_i^k(m)$ as $m$-th element \\
$\mathbf{W}_g^k \in \mathbb{R}^{M \times d}$ & Learnable feature transformation matrix \\
$\mathbf{b}_g^k \in \mathbb{R}^{M \times 1}$ & Learnable bias matrix \\
$\mathbf{z}_i^k$ & Output latent representation of corresponding gate \\
$\mathbf{z}_i^{0,k'}$ & Gate output for $k=0$ (non-treatment) in branch $k'$ \\
$\hat{y}_i^k$ & Predicted response under treatment $k$ \\
$\hat{y}_i^{0,k}$ & Predicted response under no treatment by branch $k$ \\
$\hat{y}_i^{0,*}$ & Adjusted approximate response under no treatment \\
$h^k(\cdot)$ & Tower function for $k$-th branch \\
$\mathcal{L}$ & Total training loss \\
$D_{KL}(\cdot\|\cdot)$ & Kullback-Leibler divergence \\
$\delta_k(\mathbf{x}_i)$ & Inter-day static page preference score for page $k$ \\
\bottomrule
\end{tabular}
}
\end{table}

\begin{table}[h]
\centering
\caption{KLAN-IIT Specific Notations}
{\renewcommand{\arraystretch}{1.2}
\begin{tabular}{ll}
\toprule
\textbf{Symbol} & \textbf{Description} \\
\midrule
$Q^\pi: \mathcal{S} \times \mathcal{A} \rightarrow \mathbb{R}^{K \times 1}$ & Main Q-network \\
$Q^\mu: \mathcal{S} \times \mathcal{A} \rightarrow \mathbb{R}^{K \times 1}$ & Target Q-network \\
$\theta^\pi$, $\theta^\mu$ & Parameters of main and target Q-networks \\
$Q^*(s,a)$ & Optimal Q-function satisfying Bellman optimality \\
$\mathcal{D}$ & Offline dataset for CQL training \\
$\mathcal{L}_{TD}$ & Temporal difference loss \\
$\tilde{Q}_t$ & TD target: $r(s_t, a_t) + \gamma \max_{a'} Q^{\mu}(s_{t+1}, a')$ \\
$\mathcal{L}_{Reg}$ & Conservative regularization penalty \\
$\mathcal{L}_{CQL}$ & Overall CQL objective \\
$\alpha_t > 0$ & Dynamic conservative coefficient\\
$p_k(s_t)$ & Intra-day interest score for page $k$ \\
\bottomrule
\end{tabular}
}
\end{table}

\begin{table}[h]
\centering
\caption{KLAN-AM Specific Notations}
{
\renewcommand{\arraystretch}{1.2}
\begin{tabular}{ll}
\toprule
\textbf{Symbol} & \textbf{Description} \\
\midrule
$\mathbf{c}_t$ & Contextual features at time $t$ \\
$\mathbf{v}_t$ & users' long-term behavioral statistics at time $t$ \\
$\boldsymbol{\gamma}_t = [\gamma_t^1, ..., \gamma_t^K] \in [0,1]^K$ & Context-aware weight vector \\
$\gamma_t^{k}$ & Relative importance of static preferences for page $k$ \\
$\sigma_k$ & Final navigation score for landing page $k$ \\
$k^*$ & Optimal landing page based on highest fused score \\
\bottomrule
\end{tabular}
}
\end{table}

\end{document}